%% file: machining_MLTDA.tex
\documentclass[dvipsnames]{article}

\usepackage[margin=1in]{geometry}

\usepackage{graphicx,amsmath,upgreek,bm}
\usepackage{colortbl}
\usepackage{wrapfig}
\usepackage[rgb, table]{xcolor}
\usepackage{booktabs}
\usepackage{makecell}
\usepackage{pifont}
\usepackage{todonotes}
\usepackage[toc,page]{appendix}
\usepackage{multirow}
\usepackage{algorithm}
\usepackage{rotating}
\usepackage{hyperref}
\hypersetup{
   colorlinks=true,
   linkcolor=blue,
   filecolor=magenta,
   urlcolor=blue,
   citecolor=blue,
	}
\usepackage[
  separate-uncertainty = true,
  multi-part-units = repeat
]{siunitx}
\usepackage{caption} 
\usepackage{subcaption}
\usepackage{amsmath}
\usepackage{gensymb}
\usepackage{mathtools}
\usepackage{amssymb}  
\usepackage{appendix}    
\usepackage{booktabs} 
\usepackage[para,online,flushleft]{threeparttable}
\usepackage[noend]{algpseudocode}

\usepackage{arydshln}
\makeatletter
  \renewcommand*\env@matrix[1][*\c@MaxMatrixCols c]{%
    \hskip -\arraycolsep
    \let\@ifnextchar\new@ifnextchar
  \array{#1}}
\makeatother
\makeatletter
\def\BState{\State\hskip-\ALG@thistlm}
\makeatother
\definecolor{light-gray}{gray}{0.9}

\graphicspath{{./figures/}{../figures}}

\title{Topological Feature Vectors for Chatter Detection in Turning Processes}

\author{Melih C. Yesilli\\
				Department of Mechanical Engineering\\
				Michigan State University\\
				yesillim@msu.edu
			\and
				Firas A.~Khasawneh\\
				Department of Mechanical Engineering\\
				Michigan State University\\
				khasawn3@egr.msu.edu
			\and
				Andreas Otto\\
				Institute of Physics\\
				Chemnitz University of Technology\\
				andreas.otto@physik.tu-chemnitz.de
				}
				
\date{}

\begin{document}
\maketitle

\begin{abstract}
Machining processes are most accurately described using complex dynamical systems that include nonlinearities, time delays and stochastic effects. Due to the nature of these models as well as the practical challenges which include time-varying parameters, the transition from numerical/analytical modeling of machining to the analysis of real cutting signals remains challenging. Some studies have focused on studying the time series of cutting processes using machine learning algorithms with the goal of identifying and predicting undesirable vibrations during machining referred to as chatter. 
These tools typically decompose the signal using Wavelet Packet Transforms (WPT) or Ensemble Empirical Mode Decomposition (EEMD). 
However, these methods require a significant overhead in identifying the feature vectors before a classifier can be trained. 
In this study, we present an alternative approach based on featurizing the time series of the cutting process using its topological features. 
We first embed the time series as a point cloud using Takens embedding. 
We then utilize Support Vector Machine, Logistic Regression, Random Forest and Gradient Boosting classifier combined with feature vectors derived from persistence diagrams, a tool from persistent homology, to encode chatter's distinguishing characteristics. 
We present the results for several choices of the topological feature vectors, and we compare our results to the WPT and EEMD methods using experimental turning data.  
Our results show that in two out of four cutting configurations the TDA-based features yield accuracies as high as $97\%$. 
We also show that combining B\'ezier curve approximation method and parallel computing can reduce runtime for persistence diagram computation of a single time series to less than a second thus making our approach suitable for online chatter detection.
\end{abstract}


\input{section/sec-intro}

\input{section/sec-Data_collection}
\input{section/sec-Data_processing}
\input{section/sec-Data_tagging}

\input{section/sec-TDA-background}
\input{section/sec-method}

\input{section/sec-featurization}
\input{section/sec-Results}
\input{section/sec-Conclusion}

%
\newpage
\bibliography{ML-TDA}
\input{section/sec-appendix}

\end{document}

%% file: section/sec-intro.tex
\section{Introduction}
\label{sec:intro}
Chatter is characterized by excessive vibrations that can negatively affect the surface finish or shorten the lifetime of cutting tools during machining operations such as turning and milling. 
Consequently, identification and mitigation of chatter has become a prominent research topic in recent decades. 
Some of the challenges associated with chatter identification is that it is a complex phenomenon that depends on several factors including the dynamic properties of the tool and the workpiece. 
Therefore, as these properties vary during the cutting process, the results of predictive models become invalid, thus necessitating a data-based approach for more reliable chatter detection. 
Motivated by this goal, many studies in the literature have focused on extracting chatter features from signals obtained using sensors mounted on the cutting center. 
Most of these studies are based on analyzing the spectrum of force or acceleration signals often in combination with machine learning techniques.

For example, Thaler et al. utilized short time Fourier transform with quadratic discriminant analysis to detect chatter in a band sawing process \cite{Thaler2014}. In addition,
Yesilli et al. extracted features from experimental turning signals using Fast Fourier Transform (FFT), Power Spectral Density (PSD) and the Auto-correlation Function (ACF) \cite{Yesilli2020a}. 
However, the two most common methods for analyzing cutting signals are the Wavelet Packet Transform (WPT) and the Empirical Mode Decomposition (EMD). 
Chen and Zheng used WPT to isolate the part of the signal carrying chatter markers, and they utilized it to compute both time and frequency domain features \cite{Chen2017} that can be used as feature vector for machine learning.  
Similarly, Ji et al. used EMD to decompose the acceleration signal from the tool vibration into Intrinsic Mode Functions (IMFs) \cite{Ji2018}. 
They then identified the IMFs that carry chatter information called the informative IMFs and they used them to define feature vectors for chatter identification. 
Li et al. utilizes coarse-grained entropy rate to detect chatter in turning \cite{Li2008}. In addition, Liu et al.~applied EMD to servo motor current signal to compute the energy and the kurtosis of the informative IMFs as features for chatter \cite{Liu2011}. 
They then used Support Vector Machine (SVM) algorithm to train a classifier for chatter detection. 
SVM is the most widely used machine learning algorithm for chatter classification from sensory signals. 
Some less commonly used methods include logistic regression \cite{Ding2017}, deep belief network \cite{Fu2015}, back propagation neural network \cite{Lamraoui2013}, convolutional neural networks~\cite{Cheng2019}, and Hidden Markov Model \cite{Han2016,Xie2016}.
Zuo et al. proposed an approach which is based on spiking neural networks for fault detection in bearings \cite{Zuo_2020}.
Although WPT and EEMD are widely used for detecting chatter, these methods have some limitations that preclude them from being adopted as general chatter detection tools. 
To elaborate, Yesilli et al. recently showed that identifying appropriate feature vectors using these two methods is signal-dependent and it requires skilled operators \cite{Yesilli2020}. 
Further they also showed that the accuracy of the trained classifier is highly sensitive to changes in the dynamic properties of the tool-workpiece system \cite{Yesilli2020}. In addition, to these the features obtained from signal decomposition methods does not provide extensive information about the data set. Therefore, we propose a novel approach based on Topological Data Analysis (TDA) which removes the requirement of preprocessing. This increases the level of automation of feature extraction. In addition,  we can have more information about the structure of the data using the features obtained with TDA. 

Topological Data Analysis (TDA) \cite{Ghrist2008,Carlsson2009,Edelsbrunner2009,Oudot2015} is a promising tool for generating feature vectors for chatter detection comes from a new field with many mature computational tools.
TDA, and more specifically persistent homology, provides a quantifiable way for describing the topological features in a signal \cite{Robinson2014}. 
Specifically, by embedding the sensory signal into a point cloud, it is then possible to use persistent homology to produce a multiscale summary of the topological features of the signal thus enabling the analysis of the underlying dynamical system. 
The homology classes that correspond to the embedded signal are often reported using a planar diagram that shows how long each topological feature persisted. 
The application of TDA tools to machining dynamics has only been recently explored \cite{Khasawneh2014b,Khasawneh2016,Khasawneh2018}. 
Specifically, Ref.~\cite{Khasawneh2016} and \cite{Khasawneh2014a} show that maximum persistence---a single number from the persistence diagram---can be used to ascertain the stability of simulated data from a stochastic turning model. 
Khasawneh et al. incorporated more information from the persistence diagram by extracting $5$ features including Carlsson coordinates (\cite{Adcock2016}) and the maximum persistence \cite{Khasawneh2018}, see Section \ref{sec:pd_featurization} for more details on featurizing persistence diagrams. 
The resulting feature vector in combination with SVM was used to train a chatter classifier, and it was applied to simulated deterministic and stochastic turning data with success rates as high as $97\%$ in the deterministic case.
In addition, Yesilli et al. utilized Carlsoon Coordinates and Template Functions~\cite{Perea2019} to diagnose chatter in milling simulations and show that these two featurization methods are noise robust \cite{Yesilli2019}.

However, despite the active work in the literature on featurizing persistence diagrams, all prior studies on chatter detection with TDA have utilized only a small fraction of the persistence diagram for constructing a feature vector. 
Further, these publications only studied simulated signals and no sensory signals from actual cutting tests have been tested.  
Therefore, this work aims to collect and summarize state-of-the-art featurization tools for persistence diagrams, and apply them for the first time for chatter classification using actual experimental signals obtained from an accelerometer mounted on the cutting tool during a turning process. 
The methods that we investigate for featurizing the resulting persistence diagrams and classifying chatter time series include persistence landscapes~\cite{Bubenik2015}, Carlsson coordinates~\cite{Adcock2016}, persistence images~\cite{Adams2017}, an example kernel method~\cite{Reininghaus2015}, and path signatures of persistence landscapes~\cite{Chevyrev2018}. 
Moreover, we provide the run time for each featurization method including the runtime for persistence diagram computation which constitutes the majority of the total computation time.
To reduce the runtime for persistence diagram computation, we utilize B\'ezier curve approximation method~\cite{Tsuji2019}, greedy permutation~\cite{Cavanna2015} and parallel computing.

The paper is organized as follows. 
In Section~\ref{sec:pd_featurization} we provide a brief literature survey of featurization methods for persistence diagrams. 
In Section~\ref{sec:experiment} we describe the experimental apparatus as well as how the data was processed and tagged. 
Section~\ref{sec:TDA} provides the needed background material on the tools we use from TDA. 
The method we use for featurizing the persistence diagrams and how we compute persistence diagrams are explained in Section~\ref{sec:method}. 
Section~\ref{sec:results} presents the classification results and runtime comparisons, while the concluding remarks can be found in Section~\ref{sec:Conclusion}. 

\subsection{An overview for featurization of persistence diagrams}
\label{sec:pd_featurization}
Persistence diagrams can be compared to each other using metrics such as the Wasserstein and Bottleneck distances. 
This allows creating similarity or distance matrices between different persistence diagrams that can be used for machine learning. 
For example, in previous studies on machine learning applications with TDA, these two distances were utilized as similarity measures between persistence diagrams for classification algorithms~\cite{Li2014}.  
However, working directly with persistence diagrams is difficult in part due to the non-uniqueness of geodesics which leads to non-uniqueness of the Fr\'echet mean for a collection of diagrams~\cite{Mileyko2011,Turner2014,Munch2015}. 
This necessitates using methods for extracting feature vectors that live in a Euclidean space from the persistence diagrams, which enables using traditional machine learning tools with topological features. 
The following paragraph provides a list of some of these methods. 

Adcock et~al.~describe a subring of polynomial functions on persistence diagrams and a convenient system of free generators~\cite{Adcock2016}. 
However, their construction is not continuous with respect to the bottleneck distance, and it is not applicable to the case of infinite persistence diagrams which are relevant when working with fractal structure. 
Other tools for extracting feature vectors from persistence diagrams are based on functional summaries~\cite{Berry2018}, which turn persistence diagrams into functions. 
Perhaps the most common functional summary of persistence diagrams is persistence landscapes~\cite{Bubenik2015} which represent persistence diagrams in the form of piecewise continuous functions. 
Although averages of persistence diagrams are not well-defined, persistence landscapes have well-established statistical properties~\cite{Chazal2014b} and their averages and pairwise distances are defined which enables combining them with traditional machine learning tools. 
Another functional summary of persistence diagrams is persistence images ~\cite{chepushtanova2015persistence,Adams2017}, which are closely related to Chen et al.~persistence intensity functions~\cite{chen2015statistical}. 
Persistence images are also related to prior works on the size function which predates persistence~\cite{Donatini1998,Ferri1998}. 
These images are obtained by placing Gaussians at each point in the diagram, and then utilizing a histogram of the image for featurization and machine learning.

Another featurization method is based on path signatures of persistence landscapes~\cite{Chevyrev2018}). 
Specifically, signatures for $n$-dimensional paths of bounded variation can be computed and used as features for machine learning~\cite{Chevyrev2016}. 
Therefore, by composing persistence landscapes, which can be viewed as one type of persistence path embedding, with path signatures, it is possible to extract mesh-free feature vectors from the underlying persistence diagram. 
However, path signatures do not come with stability guarantees for all input paths. 

Persistence diagrams can also be featurized using a general class of functions called template functions~\cite{Perea2019}. 
These functions are only required to be continuous and compactly supported. 
The specific realization of these functions that we work with in this paper are Chebyshev interpolating polynomials~\cite{Fox1968}. 

Kernel-based methods have also been used for featurizing persistence diagrams~\cite{Reininghaus2015,Kwitt2015,Zhao2019,Kusano2016, Kusano2017, Carriere2017c,Kusano2018}). 
However, instead of testing all of the available options, we choose the persistence scale space kernel given by~\cite{Reininghaus2015} as a representative of this group of methods. 
This kernel is defined by treating the persistence diagram as a sum of Dirac deltas at each point in the persistence diagram, and using this as the initial condition for a heat diffusion problem. 
This allows obtaining a closed form solution for this kernel.

%% file: section/sec-Data_collection.tex
\section{Experimental Setup and Signal Processing}
\label{sec:experiment}
This section describes the details of the cutting tests \ref{sec:experiment-setup}, the data collection, signal tagging \ref{sec:data_tagging}, and time series conditioning \ref{sec:Data_Processing}. 
\subsection{Cutting tests}
\label{sec:experiment-setup}
Figure~\ref{fig:experimental_setup} shows the experimental setup which features a Clasuing-Gamet $33$ cm ($13$ inch) engine lathe instrumented with three accelerometers.  
Two PCB 352B10 uniaxial accelerometers were used to measure the $x$ and $y$ accelerations of the boring rod, see Fig.~\ref{fig:experimental_setup}. These uniaxial accelerometers have frequency range between 2 and $10000$ Hz and sensivity of 1.02 mV/(m/$s^{2}$).
They were mounted $3.81$ cm ($1.5$ inch) away from the cutting tool to keep them safe from the cutting debris.
Further, a PCB 356B11 triaxial accelerometer was attached to the tool holder with superglue. Triaxial accelerometer has frequency range between $2$ to $10000$ Hz for $y$ and $z$ axes and it has range of $2$ to $7000$ Hz for $x$ axis. It has also sensitivity of 1.02 mV/(m/$s^{2}$).
The acceleration signals were collected with a sampling rate of $160$ kHz using 
Matlab's data acquisition toolbox and a NI USB-6356 data acquisition box, which has 8 analog inputs with 1.25MS/s/ch sampling rate and 16-bit resolution, see Section~\ref{sec:Data_Processing} for more information on the applied signal processing.  

The used cutting tool is a $0.04$ cm ($0.015$ inch) radius Titanium nitride coated insert which was attached to an S10R-SCLCR3S boring bar, where the latter is part of the Grizzly T10439 carbide insert boring bar set. This boring bar set uses the $80^{\circ}$ diamond cutting inserts and has negative $7^{\circ}$ end and side cutting angle. 
6061 aluminum cylindrical workpiece with Brinell hardness of 95 is used in the cutting tests. 

The cutting tests were performed by varying the depth of cut and the cutting speed while holding the stiffness of the boring rod constant. Some of the cutting depths are 0.0127 cm, 0.0508 cm and 0.0635 cm (0.005 inch, 0.02 inch and 0.025 inch). Each vibration signal is labeled with its cutting speed and depth of cut and available in the Mendeley repository \cite{Khasawneh2019}. The feed rate for all cutting tests was kept constant and it is 0.00508 cm/rev (0.002 inch/rev).
Data was collected from four sets of cutting configurations where each configuration corresponds to a different boring rod stiffness/eigenfrequency. Experimental data is available online in raw and processed format \cite{Khasawneh2019}.
The stiffness of the rod was varied by changing the overhang length of the tool, which is the distance between the heel of the boring rod and the back surface of the tool holder. 
Four different overhang lengths were considered: $5.08$ cm ($2$ inch), $6.35$ cm ($2.5$ inch), $8.89$ cm ($3.5$ inch) and $11.43$ cm ($4.5$ inch).  
Increasing the overhang length leads to a stiffer rod and to higher chatter frequencies since the latter are close to the eigenfrequency of the rod. 
Similarly, smaller overhang length leads to a more compliant system, and to lower chatter frequencies. 

\begin{figure}[h!]
\centering
\captionsetup{justification=centering}
\includegraphics[width=0.6\textwidth,height=.75\textheight,keepaspectratio]{experimental_setup_fig.png}
\caption{The experimental setup used for collecting cutting data.}
\label{fig:experimental_setup}
\end{figure}

%% file: section/sec-Data_processing.tex
\subsection{Signal Processing}
\label{sec:Data_Processing}
The experimental data was sampled at the rate of $160$ kHz, which is an oversampling by a factor of $16$ of the maximum accelerometer measuring frequency of $10$kHz. 
Since no in-line analog filter was used, this oversampling is necessary to avoid aliasing effects by utilizing a digital filter before downsampling the signal to $10$kHz. 
The filter we used is Butterworth low pass filter with order $100$ from Matlab's signal processing toolbox. 
The resulting filtered and downsampled signals were the ones used in tagging the data as described in Section~\ref{sec:data_tagging}. 
 

%% file: section/sec-Data_tagging.tex
\subsection{Data Tagging}
\label{sec:data_tagging}
Upon examining the acceleration signals, we found that the $x$-axis signals from the tri-axial accelerometer (see Fig.~\ref{fig:experimental_setup}) contained the best Signal-to-Noise Ratio (SNR). 
The other acceleration signals contained redundant information albeit with lower SNR; therefore, only the $x$-axis signal from the tri-axial accelerometer was used for tagging the data. 
Both time and frequency domain information were simultaneously considered when tagging the data. 
Specifically, in the time domain, we separated the different parts of the raw time series according to amplitude, i.e., portions with large amplitude, low amplitude, and with sudden increase in amplitude. 
In the frequency domain, frequencies under $5$ kHz in the downsampled signal were taken into account. 

\begin{figure}[h]
\centering
\includegraphics[width=0.75\textwidth,height=.80\textheight,keepaspectratio]{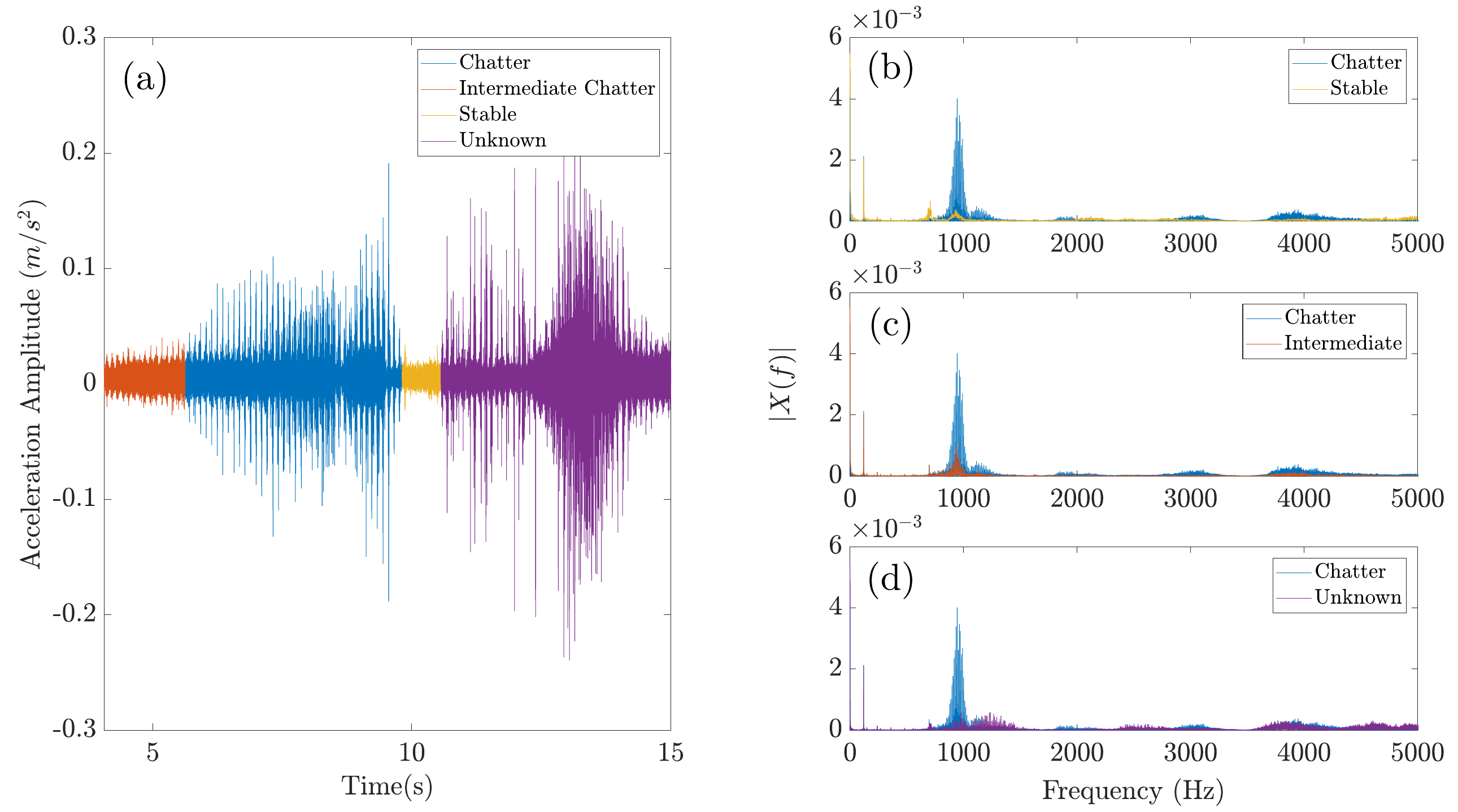}
\caption{(a) An example of a tagged time series, and a plot of the spectrum of the chatter regime superimposed onto the spectrum of: (b) chatter-free cutting, (c) intermediate chatter, and (d) unknown case. The signal is for turning with a overhang length of $5.08$ cm ($2$ inch), $570$ rpm, and $0.00508$ cm depth of cut($0.002$ inch).}
\label{fig:raw_data_fft_downsampled_2inch_570_002}
\end{figure}

We established specific criteria to label the raw data which we describe below. 
We also provide the number of tagged time series for each overhang case in Table \ref{tab:chatter_case_number}.
Figure~\ref{fig:raw_data_fft_downsampled_2inch_570_002}a shows that during the same cutting test it was possible for different parts of the time series to be tagged differently. 
A signal was tagged chatter-free (stable) if it has a low amplitude in both time and frequency domains. 
These signals have the highest peaks at the spindle rotation frequencies in the Fourier spectrum \cite{Insperger2008}, as shown in Fig.~\ref{fig:raw_data_fft_downsampled_2inch_570_002}b. 
On the other hand, we characterized intermediate chatter signals as having low amplitude in the time domain but high amplitude in the frequency domain, see Fig.~\ref{fig:raw_data_fft_downsampled_2inch_570_002}c. 
In contrast, chatter signals have large amplitude in both time and frequency domains, as shown in Fig.~\ref{fig:raw_data_fft_downsampled_2inch_570_002}a and the spectrum superimposed on Figs.~\ref{fig:raw_data_fft_downsampled_2inch_570_002}b--\ref{fig:raw_data_fft_downsampled_2inch_570_002}d. 
For intermediate chatter and chatter signals, the highest peaks are around the chatter frequencies in the frequency domain. 
Any case that did not fit with any of the above categories was labeled as unknown. 
Figure~\ref{fig:raw_data_fft_downsampled_2inch_570_002}d shows that these cases are generally composed of time series that have sudden increase and decrease in amplitude in the time domain combined with low amplitude in the frequency domain. 

\begin{figure}[h]
\centering
\includegraphics[width=0.75\textwidth,height=.75\textheight,keepaspectratio]{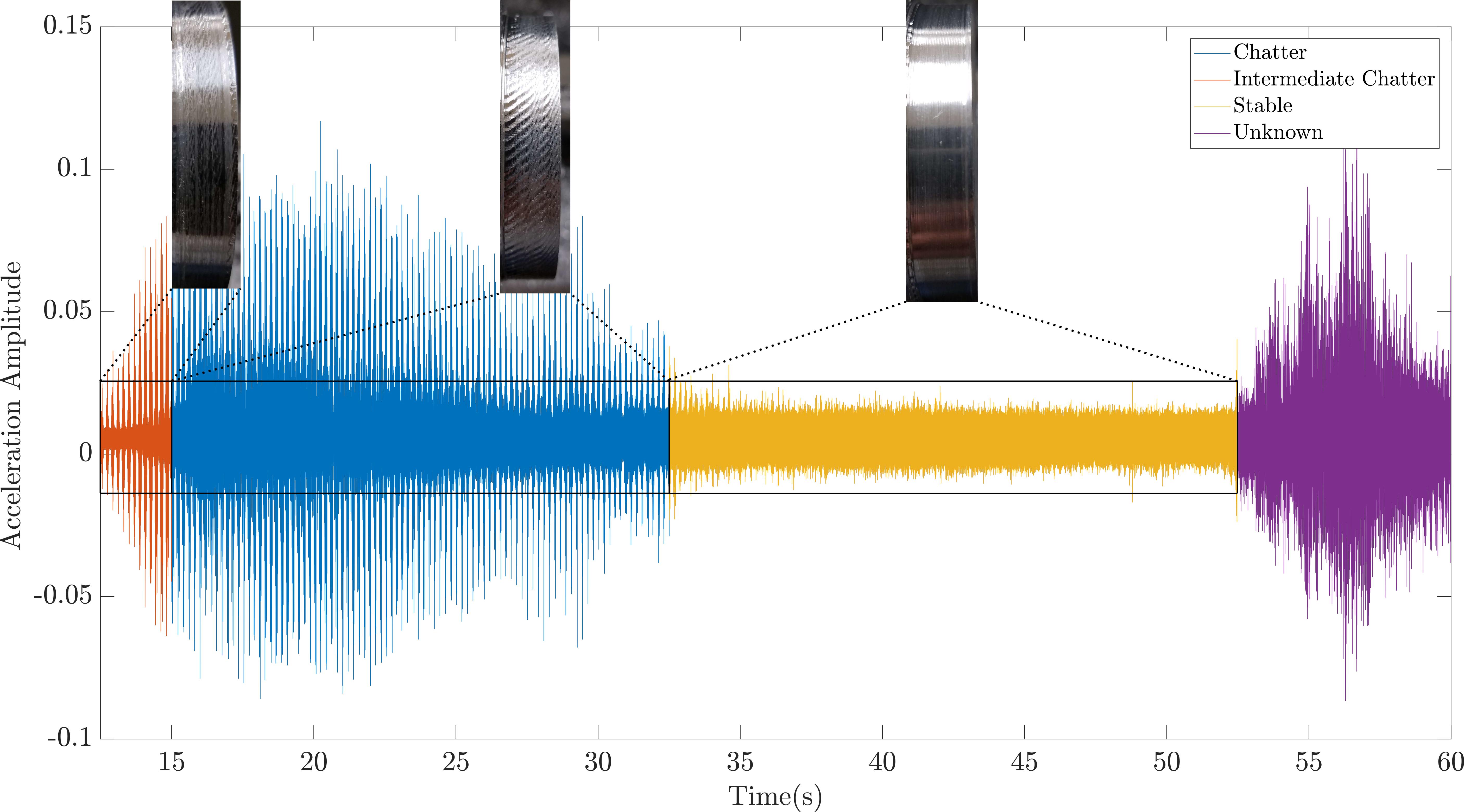}
\caption{Tagged raw data and corresponding cut surfaces for 5.08 cm (2 inch) 320 rpm and 0.0127 cm depth of cut (0.005 inch) case.}
\label{fig:2inch_raw_data_cut_surface}
\end{figure}

The data labeling was verified by spot-checking the tagged data against pictures taken during the cutting tests.  
Figure~\ref{fig:2inch_raw_data_cut_surface} shows one such comparison for a sample tagged signal. 
The figure shows that the surface finish characteristics agree with the assigned tagging, thus confirming the validity of the labeling.

\begin{table}
\centering
	\begin{tabular}{ccccc}
		\makecell{Overhang length\\ (cm (inch))} & \makecell{ Stable} & \makecell{Mild chatter} & \makecell{Chatter }  & \makecell{Total} \\
		\toprule 
		5.08 (2)	  & 19 	& 8   & 12 & 39 \\
		6.35 (2.5)	& 9 	& 4   & 3  & 16 \\
		8.89 (3.5)  & 9 	& 3   & 2  & 14 \\
		11.43 (4.5) & 13 	& 4   & 5  & 22 \\
		\bottomrule
	\end{tabular}
	\centering
	\caption{The number of tagged datasets for each overhang case.}
	\label{tab:chatter_case_number}
\end{table}

%% file: section/sec-TDA-background.tex
\section{Topological Data Analysis}
\label{sec:TDA} 
Topological Data Analysis (TDA) extracts information by investigating the shape of the data. 
We study persistent homology which is a powerful tool of Topological Data Analysis (TDA) to extract features from the persistence diagrams and use them in supervised machine learning algorithms. Experimental data is embedded using Takens embedding theorem \cite{Takens1981} and 1-D persistent homology is investigated for feature matrix generation. 
In this section, we briefly explain persistent homology, and one can refer to \cite{Ghrist2008,Carlsson2009,Edelsbrunner2009,Oudot2015,Munkres2018,Munch2017} for detailed information about TDA and persistent homology. 

Persistent homology provides a compact tool for studying the topology of data embedded in a Euclidean space which is often called a point cloud. 
The resulting shape information is represented by a two dimensional plot called the persistence diagram. 
We can obtain a persistence diagram for different shape characteristics of interest. 
For instance, if we are interested in the connectivity of the points in the point cloud, then we can represent that information in a $0$-dimensional ($0$-D) persistence diagram. 
Alternatively, if we are interested in loops in the point cloud, then we compute the $1$-D persistence diagram. 
For voids, we go to the $2$-D persistence diagram and so on. 
In this study we only extracted features from the $0$-D and $1$-D persistence, so in the following we will introduce the basic idea for obtaining the $1$-D persistence, i.e., for representing loops that emerge and disappear as the point cloud is thickened. 
The process for obtaining the $0$-D persistence is similar, only instead of considering loops, we would need to track the connectivity of the points as the point cloud is uniformly thickened.    
 
\begin{figure}[h]
\centering
\includegraphics[width=0.85\textwidth,height=.75\textheight,keepaspectratio]{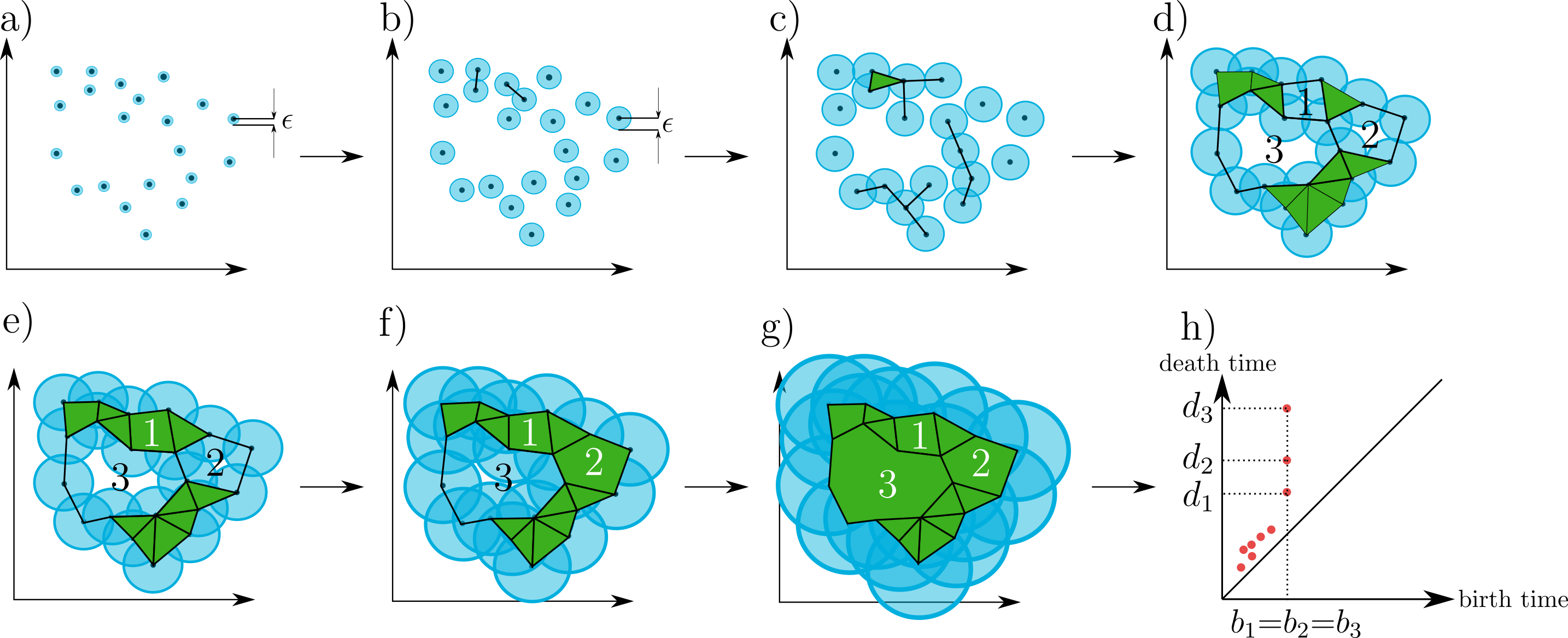}
\caption{Generation of persistence diagrams using The Rips Complex.}
\label{fig:Rich_Complex}
\end{figure}

Consider the point cloud shown in Fig.~\ref{fig:Rich_Complex}a.  
Then, we start to thicken the point cloud, i.e., expand disks with radius $\epsilon$ around each data point.
As $\epsilon$ is increased, disks can start to intersect. 
The intersection of two disks forms an \textit{edge} as shown by the two edges in Fig.~\ref{fig:Rich_Complex}b. 
Increasing $\epsilon$ further can lead to three disks intersecting thus forming a triangle which we fill in as shown by the nine triangles in Fig.~\ref{fig:Rich_Complex}c. 
At some values of $\epsilon$, some disk intersections will lead to \textit{cycles}.  
The time (here the $\epsilon$ value) at which a cycle appears is called the birth time of the cycle. 
Figure~\ref{fig:Rich_Complex}d shows three example cycles numbered 1,2 and 3 with birth time  $b_{1}=b_{2}=b_{3}$.  

As we continue thickening the disks in the point cloud, more disks will intersect leading to more triangles filling in, and at some point, some cycles may fill in.  
The time at which a cycle disappears is called its death time. 
For example, Figs.~\ref{fig:Rich_Complex}e--g show the death times of cycles $1$--$3$, respectively. 
The information about the birth and death of cycles is succinctly summarized in a persistence diagram. 
In this diagram, each point corresponds to the paired birth and death times of a cycle. 
For example, Fig.~\ref{fig:Rich_Complex}h shows the tuples $(b_1,d_1)$, $(b_2, d_2)$, and $(b_3, d_3)$ corresponding to the birth and death times of the cycles $1$--$3$ shown in Fig.~\ref{fig:Rich_Complex}d. 
The cycle that persists the longest is characterized by the highest point above the diagonal, and is called the maximum persistence. 
In this example, cycle $3$ leads to the maximum persistence in the persistence diagram.   
  
\subsection{Simplicial complexes}
\label{sec:SimplicialComplexes}
\begin{figure}[h!]
\centering
\includegraphics[width=0.75\textwidth,height=.75\textheight,keepaspectratio]{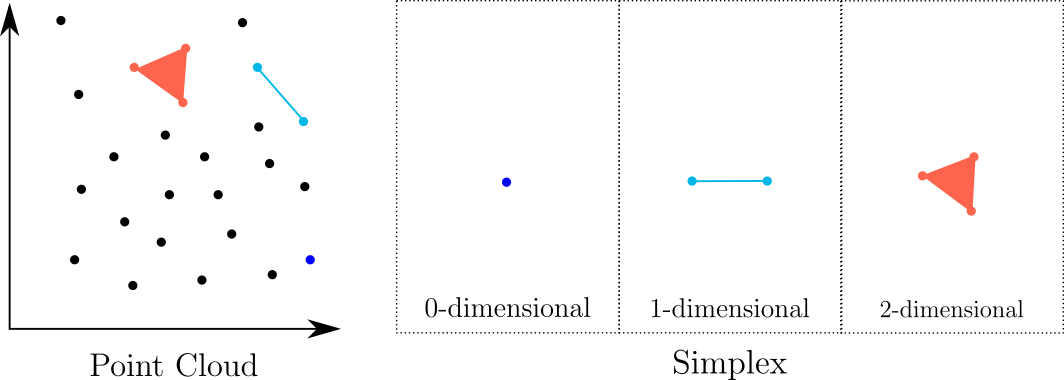}
\caption{Formation of simplicial complexes from point cloud.}
\label{fig:simplicial_complex}
\end{figure}

Let $\{u_{0},\ldots,u_{k}\} \in \mathbb{R}^{d}$ be a  set of data points, and the vectors defined between these data points ($u_{1}$-$u_{0}$,$u_{2}$-$u_{0}$,\ldots,$u_{k}$-$u_{0}$) are linearly independent.
A geometric \textit{k-}simplex, $\sigma$ 
is a set of all points in $\mathbb{R}^{d}$ such that
$\sum_{j=0}^{k} \lambda_{j}u_{j}$ where $\sum_{j=0}^{n} \lambda_{j} =1$ and $\lambda_{j}\geq 0$ for all j. 
Figure~\ref{fig:simplicial_complex} provides illustrations for $0,1,$ and $2-$simplex. 
Each data point on a point cloud is represented as 0-dimensional simplex and they are called vertices. 
When two vertices are connected, an edge is formed and it is 1-dimensional simplex. 
Connection of three vertices will form $2-$simplex which is a triangle.
Simplicies spanned by any subset of $u_{0},\ldots,u_{k}$ are the faces of $\sigma$.
In general, $n-$simplex contains $n+1$ vertices, and the set of these simplices, are called geometric simplicial complexes, $K$, if the following two conditions are satisfied \cite{Munkres2018}: 
1) If $\sigma \in K$, then faces of $\sigma$ are also in $K$, 
2) If two $n-$simplex, $\sigma_{1}$ and $\sigma_{2}$ are in $K$, then the intersection of them is either common face or empty.
The dimension of the simplicial complex, $K$ is equal to the largest dimension of its simplices. 

\subsection{Persistent Homology}
\label{sec:persistent_homology}
The simplicial complex  $K$ is used to compute homology $H_{n}(K)$ in different dimension to identify the shape of the data. 
0 dimensional homology, $H_{0}(K)$ represents connected components and one dimensional homology, $H_{1}(K)$ represents loops, while two dimensional homology $H_{2}(K)$ represents voids.
In persistent homology, the simplicial complex $K$ is not fixed and it is varying over time.

We start expanding disks centered at data points of point cloud and let $\epsilon$ be the radius of these disks. 
As $\epsilon$ increases, n dimensional simplicies are formed as shown in Fig.~\ref{fig:Rich_Complex}. Intersection of two disks forms an edge (1-simplex), while a triangle (2-simplex) is formed when three disk intersect with each other (see Fig.~\ref{fig:simplicial_complex}).
Each $\epsilon$ will result in different simplicial complexes, and they can be approximated using filtration functions. 
In this study, we used a Python package that employs Rips complex whose definition is given as
\begin{equation*}
R_{\epsilon}(K,d) = \{ \sigma \subset K | \underset{(x,y)\in \sigma}{\mathrm{max}} d(x,y)< \epsilon\},
\end{equation*} 
where $d$ is the distance between vertices of simplicial complex $\sigma$. Let \{$\epsilon_{1}<\epsilon_{2}<\ldots<\epsilon_{m}$ \} be the set of varying radius of the disk. These radii form Rips complexes such that 
\begin{equation*}
R_{1}\subseteq R_{2}\subseteq \ldots \subseteq R_{m}
\end{equation*}
where $R_{j} = R_{\epsilon_{j}}(K,d)$. Then, we can choose a specific dimension $n$ to identify the shape of the data along the simplicial complexes $R_{j}$. 
For instance, if a loop is seen first in $R_{i}$, this is called birth time ($b=\epsilon_{i}$). 
When it disappears in $R_{j}$, this is denoted as death time ($d=\epsilon_{j}$, where $d>b$). 
That allows us to generate persistence diagrams, $D$ for a given point cloud and selected persistent homology dimension. 
Figure~\ref{fig:Rich_Complex}h provides an example of a persistence diagram. 
The horizontal axis represents the birth time, while the vertical axis is for the death time. 
All points on the diagram is on the upper diagonal and the points with larger lifetime $(d-b)$ is far away from the diagonal point, and these points includes major information about the shape and the structure of the data. 
In this study, we mostly focused on the one dimensional persistent homology to extract features.

%% file: section/sec-method.tex
\section{Method}
\label{sec:method}
\begin{figure}[h!]
\centering
\includegraphics[width=1\textwidth]{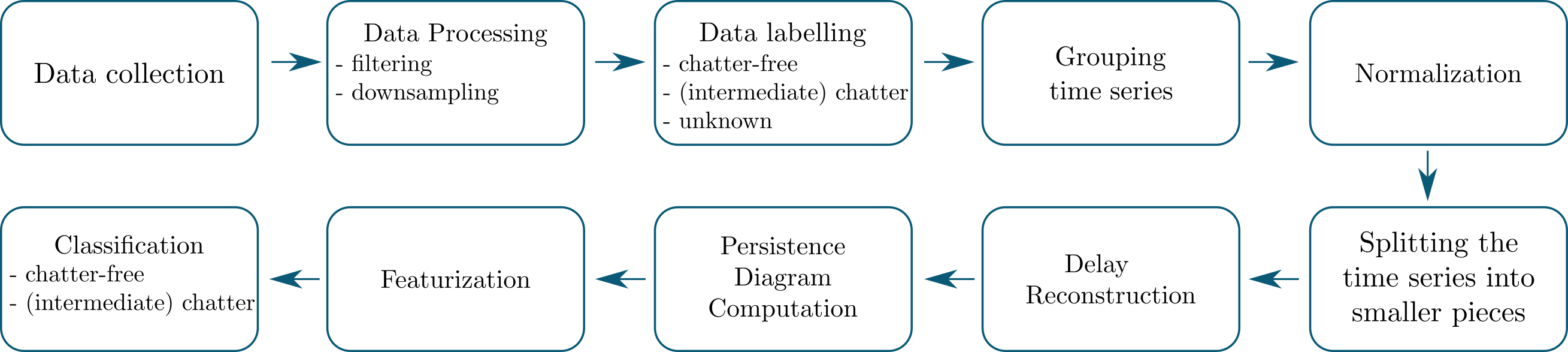}
\caption{Pipeline for feature extraction using topological features of data.}
\label{fig:process_PD}
\end{figure}

The method that we develop for chatter detection using topological features can be summarized using Fig.~\ref{fig:process_PD}. 
The parts of the pipeline related to data collection, processing, and labeling were described in Sections~\ref{sec:experiment-setup}--\ref{sec:data_tagging}, respectively. 
In this section, we describe the rest of the steps shown in Fig.~\ref{fig:process_PD}.  

Recall that the cutting tests are composed of four different overhang configurations, and that each configuration includes a different number of time series that correspond to different labels, rotational speeds, and depths of cut. 
Therefore, the time series are grouped with respect to these three parameters and they are normalized to have zero mean and unit variance. 
This normalization reduces the effect of large feature values on smaller ones (\cite{Theodoridis2009}). 

The next step is to split long time series into smaller pieces to reduce the computation time needed for finding the delay reconstruction parameters (see Section \ref{sec:delay_recon}), and for obtaining the persistence diagrams (see Section \ref{sec:persistent_homology}). 
Upon finding the appropriate embedding parameters, the data is embedded using delay reconstruction, also known as Takens' delay embedding, see Section \ref{sec:delay_recon}.  
The resulting point cloud is then used to compute the corresponding persistence diagrams using two different approaches: 1)traditional way where persistence diagrams are computed with \href{https://ripser.scikit-tda.org/}{Ripser} package for Python, 2) approximating to point cloud with Bezier curves and computing persistence diagrams using line segments generated with B\'ezier curves (\cite{Tsuji2019}).
Second method is a different approach introduced in \cite{Tsuji2019}, and it uses B\'ezier curves to approximate  to reduce time to compute persistence diagrams.
Section \ref{sec:per_diag_comp} explains both approaches for persistence diagram computation and sections \ref{sec:persistence_landscapes}--\ref{sec:persistence_paths} describe different methods in the literature for featurizing the resulting diagrams to obtain a feature vector in Euclidean space that can be used with existing machine learning tools such as SVM.  
We mainly featurize and use the resulting $1$-dimensional persistent homology $H_{1}$ except when we use the template function approach where we also utilize the $0$-dimensional persistence $H_{0}$ (\cite{Perea2019}). 

\subsection{Persistence Diagram Computation}
\label{sec:per_diag_comp}
\begin{figure}[h]
\centering
\includegraphics[width=0.95\textwidth]{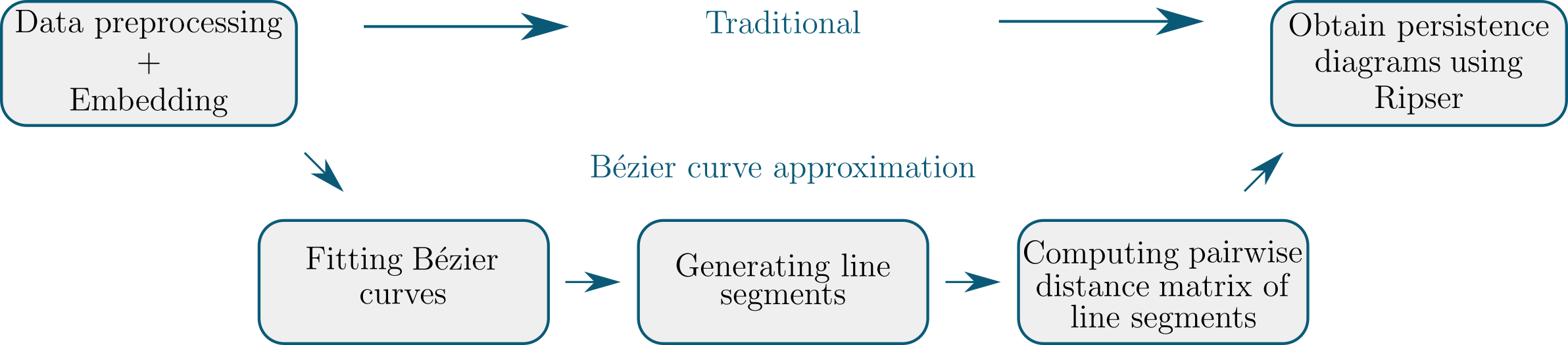}
\caption{Persistence diagram computation steps.}
\label{fig:PD_Comp_Steps}
\end{figure}

In this section, we briefly explain how to embed a time series to higher dimension and to compute persistence diagrams using B\'ezier curve approximation method.
The steps for persistence diagram computation are summarized as shown in Fig.~\ref{fig:PD_Comp_Steps}.

\subsubsection{Delay Reconstruction}
\label{sec:delay_recon}

Takens' theorem lays a theoretical framework for studying deterministic dynamical systems (\cite{Takens1981}).  
It states that in general we can obtain an embedding of the attractor of a deterministic dynamical system from a one-dimensional recording of a corresponding trajectory.  
This embedding is a smooth map $\Psi: M\to N$ between the manifolds $M$ and $N$ that diffeomorphically maps $M$ to $N$.  

Specifically, assume we have an observation function $\beta(\mathbf{x}):M  \to \mathbb{R}$, where for any time $t \in \mathbb{R}$ the point $\mathbf{x}$ lies on an $m$-dimensional manifold $M \subseteq \mathbb{R}^d$.  
While in practice we do not have the the flow of the system for an time $t \in \mathbb{R}$ given by $\phi^t(\mathbf{x}): M \times \mathbb{R} \to M$, the observation function implicitly captures the time evolution information according to $\beta(\phi^t(x))$, typically in the form of the one-dimensional, discrete and equi-spaced time series $\{\beta_n\}_{n \in \mathbb{N}}$.  

Takens' theorem states that by choosing an embedding dimension $d \geq 2m+1$, where $m$ is the dimension of a compact manifold $M$, and a time lag $\tau > 0$, then the map $\Phi_{\phi,\beta}: M \to \mathbb{R}^d$ given by
\begin{align*}
\Phi_{\phi, \beta} &= (\beta(\mathbf{x}), \beta(\phi(\mathbf{x})), \ldots, \beta(\phi^{d-1}(\mathbf{x}))) \\
                   &= (\beta(\mathbf{x}_t), \beta(\mathbf{x}_{t+\tau}, \beta(\mathbf{x}_{t+2\tau}, \ldots, \beta(\mathbf{x}_{t+(d-1)\tau}))),
\end{align*}
is an embedding of $M$, where $\phi^{d-1}$ is the composition of $\phi$ $d-1$ times and $\mathbf{x}_t$ is the value of $\mathbf{x}$ at time $t$.  

For noise-free data of infinite precision any time lag $\tau$ can be used; however, in practice, the choice of $\tau$ can influence the resulting embedding. 
In this paper $\tau$ was found by using the method of Least Median of Squares (LMS) (\cite{Rousseeuw1984}) combined with the magnitude of the Fast Fourier Transform (FFT) of the signal. 
Specifically, we obtain the FFT spectrum and we identify the maximum significant frequency in the signal using LMS.  
We then use Nyquist's sampling criterion to choose the delay value according to the inequality described in \cite{Melosik2016}.  
This approach yielded reasonable delay values in comparison to the standard mutual information function approach (\cite{Fraser1986}) where the mutual information function is plotted for several values of $\tau$ and the first dip in the plot indicates the $\tau$ value to use.  
This is because (1) the mutual information function is not guaranteed to have a minimum, thus leading to a failed selection of $\tau$ and, (2) the identification of the first \textit{true} dip, if it exists, is not easy to automate especially in non-smooth plots.  

The embedding dimension $d \in \mathbb{N}$ is computed by using the False Nearest Neighbor (FNN) approach (\cite{Kennel1992a,Abarbanel1994}). 
In the FNN approach, the delay reconstruction is applied to the time series using increasing dimensions.  
The distances between neighboring points in one dimension are re-computed when the points are embedded into the next higher dimension.  
By keeping track of the percent of points that appear to be neighbors in a low dimension but are farther apart in a higher dimension (termed false neighbors), it is possible to identify a threshold that indicates that the attractor has been sufficiently unfolded.  
Applying FNN to all of the time series yielded the values in the range $d \in \{1, 2, \ldots, 10\}$, depending on the time series being reconstructed.   
Upon identifying $\tau$ and $d$ for each time series, delay reconstruction was used to embed the signal into a point cloud $P \subseteq \mathbb{R}^d$.  
The shape of the resulting point cloud was then quantified using persistence, as described in Section \ref{sec:TDA}. 
We study five different methods to extract features from the resulting persistence diagrams as we show in Sections~\ref{sec:persistence_landscapes}--\ref{sec:persistence_paths}. 

\subsubsection{Persistence Diagram Computation with B\'ezier Curve Approximation}
\label{sec:Bezier_Approx}
B\'ezier curves are introduced by Pierre B\'ezier and it has been used in computer aided design softwares and path planning applications for robots (\cite{Tharwat2018,Elhoseny2018,Choi2008,Hwang2003}). Recently,
\cite{Tsuji2019} utilized B\'ezier curves to speed up the persistence diagram computation.
In this approach, the first step is to divide point cloud into groups. 
User should define number of samples per group ($spg$). 
Then, the B\'ezier curves is fit to each group individually.
Second step is to generate line segments using these curves, and the number of line segments per curve ($r$) is also selected by user. 
Pairwise distance matrix between the line segments is computed and it is given to $Ripser$ as input, and it will provide the persistence diagrams as output based on selected maximum number of homology dimension.
This section explains two of three main steps, which are fitting a bezier curve and computing the distance matrix between the line segments, and the effect of $spg$ and $r$ on approximation of persistence diagrams.

\begin{figure}
\centering
\includegraphics[width=1\textwidth]{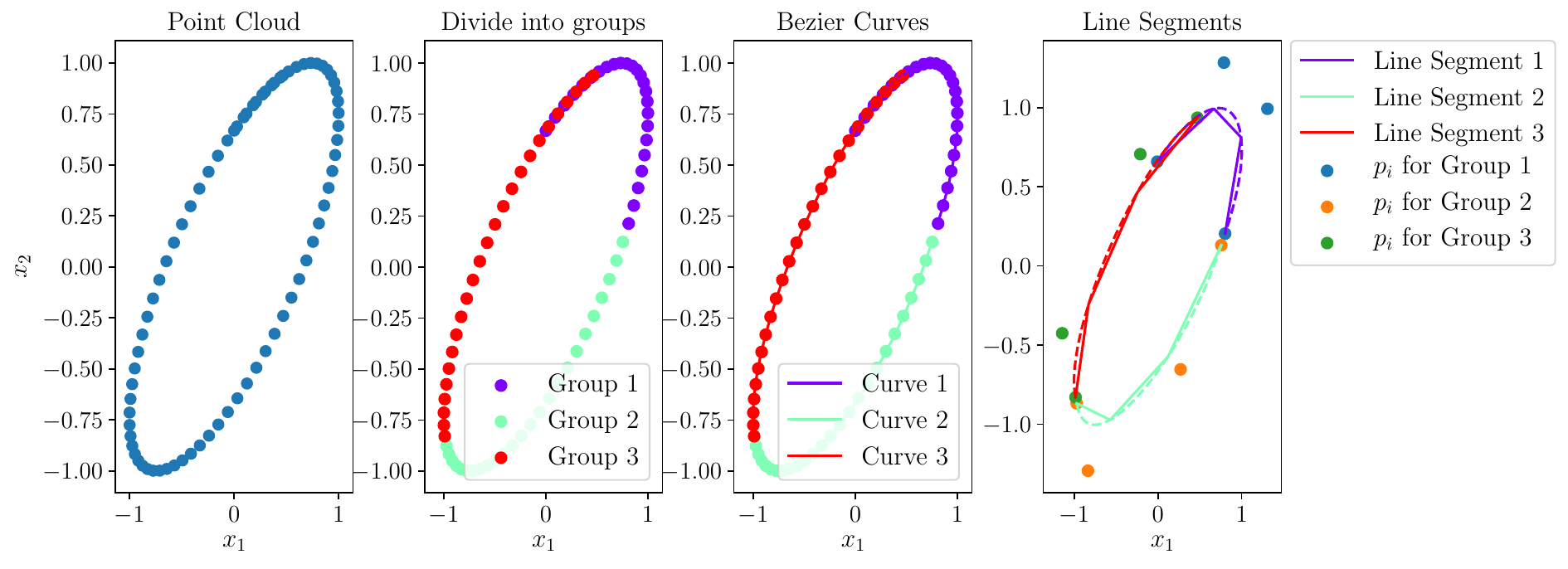}
\caption{Illustration showing B\'ezier curve fit and generation of line segments.}
\label{fig:Bezier_example}
\end{figure}

\textbf{Fitting a Bezier Curve:} In this implementation, we use cubic B\'ezier curve expression defined as 
\begin{equation*}
p(t)=\sum_{i=0}^{3}(1-t)^{3-i}t^{i}p_{i} \quad 0<t<1,
\end{equation*}
where $t$ is the parametrization variable and $p_{i}$ is called the control points of the curve. 
For cubic B\'ezier curves, we totally have four different control points.
The first one and the last one should match with the first and last point of the group of samples which we fit B\'ezier curve, respectively.
Solution for these control points is obtained by using least squares error method and the expression is given as
\begin{equation*}
\centering
L(p_{0},p_{1},p_{2},p_{3})=\sum_{i=1}^{l}||p(t_{i})-x_{i}||^{2},
\end{equation*}
where $x_{i}$ is the data points of point cloud.
Solution of $\nabla L=0$ provides the control points. The expression for $\nabla L=0$ can be rewritten such that
\begin{gather*}
\sum_{k=1}^{l}2\Bigg(\sum_{i=0}^{3} \binom{3}{i}(1-t_{k})^{3-i}t_{k}^{i}p_{i}-x_{k}\Bigg)\Bigg(\frac{\partial p(t_{k})}{\partial p_{i}}\Bigg)=0 \\
\resizebox{0.48\textwidth}{!}{$\frac{\partial p(t_{k})}{\partial p_{i}} = (1-t_{k})^{3}\hat{e}_{1}+3(1-t_{k})^{2}t_{k}\hat{e}_{2}+3(1-t_{k})t_{k}^{2}\hat{e}_{3}+t_{k}^{3}\hat{e}_{4}=0$},
\end{gather*}
where $t_{k}$ represents varying parametrization variable along a B\'ezier curve.
The above expression can also be written in matrix form as
\begin{equation*}
\Bigg(\sum_{k=1}^{l}A_{k}\Bigg)p=\sum_{k=1}^{l}b_{k} \quad \text{or} \quad Ap=b.
\end{equation*}
The equations for $A_{k}$, $p$ and $b_{k}$ are defined as
 
\begin{gather*}
\small
A_{k}=\begin{bmatrix}
(1-t_{k})^{6}       & 3(1-t_{k})^{5}t_{k}     & 3(1-t_{k})^{4}t_{k}^{2} & (1-t_{k})^{3}t_{k}^{3} \\
3(1-t_{k})^{5}t_{k}     & 9(1-t_{k})^{4}t_{k}^{2} & 9(1-t_{k})^{3}t_{k}^{3} & 3(1-t_{k})^{2}t_{k}^{4}\\
3(1-t_{k})^{4}t_{k}^{2} & 9(1-t_{k})^{3}t_{k}^{3} & 9(1-t_{k})^{2}t_{k}^{4} & 3(1-t_{k})t_{k}^{5}    \\ 
(1-t_{k})^{3}t_{k}^{3}  & 3(1-t_{k})^{2}t_{k}^{4} & 3(1-t_{k})t_{k}^{5}     & t_{k}^{6}          \\
\end{bmatrix} \\ p=
\begin{bmatrix}
p_{0}^{1} & p_{0}^{2} &\ldots &p_{0}^{d}\\
p_{1}^{1} & p_{1}^{2} &\ldots &p_{1}^{d}\\
p_{2}^{1} & p_{2}^{2} &\ldots &p_{2}^{d}\\
p_{3}^{1} & p_{3}^{2} &\ldots &p_{3}^{d}\\
\end{bmatrix}
\quad
b_{k}=\begin{bmatrix}
(1-t_{k})^{3}\\
3(1-t_{k})^{2}t_{k}\\
3(1-t_{k})t_{k}^{2}\\
t_{k}^{3}\\
\end{bmatrix}
\begin{bmatrix}
x_{k}^{1}\\
x_{k}^{2}\\
\vdots\\
x_{k}^{d}
\end{bmatrix}^{T},
\end{gather*}
where $d$ is the dimension of the data set.
To illustrate the B\'ezier curve fitting, we embed a sinusoidal signal to dimension two. Then, we divide the samples into  groups and fit bezier curves. Figure~\ref{fig:Bezier_example} represents the control points and fitted lines on the groups.

\textbf{Computing pairwise distance matrix:} B\'ezier curves are split into intervals, and the number of intervals ($r$) are selected by the user. The end points of the intervals are connected to each other and line segments are generated. The next step is to compute distance matrix between the line segments. Lets assume that $\overrightarrow{l_{0}l_{1}}$ and $\overrightarrow{m_{0}m_{1}}$ represent two lines. The distance between these two lines is defined as follows
\begin{equation*}
d(\overrightarrow{l_{0}l_{1}},\overrightarrow{m_{0}m_{1}})= \min_{l \in \overrightarrow{l_{0}l_{1}}, m \in \overrightarrow{m_{0}m_{1}}} d(l,m),
\end{equation*}
where $l_{0}$, $l_{1}$, $m_{0}$ and $m_{1}$ are the end points of the two lines.
The distance is computed by minimizing the function given as
\begin{equation*}
f(l,m)=d(l(s),m(t))^{2}=||l(s)-m(t)||^{2},
\end{equation*}
where $s$ and $t$ are parametrization variables. \cite{Tsuji2019} solved this problem using gradient descent algorithm, however we employ the simplicial homology global optimisation (SHGO) (\cite{Endres2018}). After finding the parameters making function $f$ minumum, we use them values to compute the distance between two lines.
This is repeated for all combinations between line segments. 
We only compute the upper diagonal of the distance matrix, since it is symmetric.
Then, it is given to $Ripser$ package as input to obtain persistence diagrams.

\begin{figure}[h]
\centering
\includegraphics[width=1\textwidth]{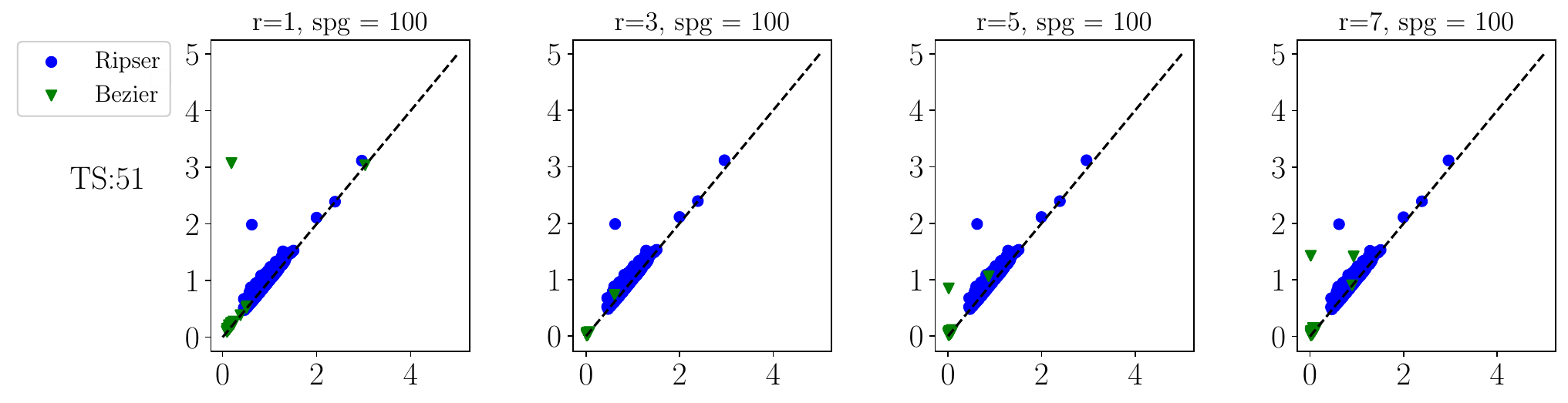}
\caption{Comparison of persistence diagrams computed with B\'ezier approximation for varying $r$ to true diagrams obtained with $Ripser$. All diagrams belong to $51^{th}$ time series of the 8.89 cm (3.5 inch) case and they are computed in $H_{1}$.}
\label{fig:Comparison_of_diagrams}
\end{figure}

\begin{figure}[h!]
\centering
\includegraphics[width=0.5\textwidth]{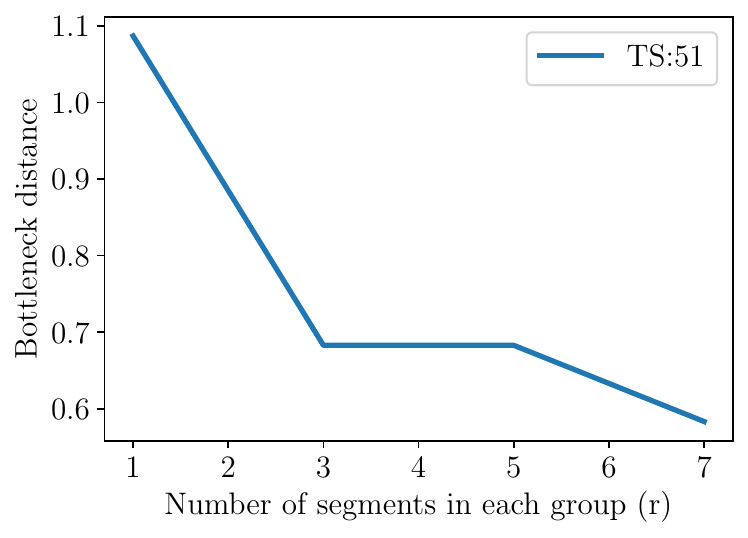}
\caption{The Bottleneck distance between the diagram with blue color in Fig.~\ref{fig:Comparison_of_diagrams} and the approximated diagrams with green color in Fig.~\ref{fig:Comparison_of_diagrams}.}
\label{fig:distances_between_pd}
\end{figure}

\textbf{Effect of parameter selection:} The parameters sample per group ($spg$) and the number of line segments ($r$) define how well we approximate to persistence diagrams of a time series. We choose a time series belong to 8.89 cm (3.5 inch) overhang distance, and we computed persistence diagrams of that time series with $r=1,3,5,7$ and $spg=100$. In Fig.~\ref{fig:Comparison_of_diagrams}, we plot both diagrams computed with $Ripser$ in blue and approximated diagrams in green color. All diagrams represents the first dimensional persistence ($H_{1}$) and they belong $50^{th}$ time series in the 8.89 cm (3.5 inch) case. Figure~\ref{fig:Comparison_of_diagrams} shows that approximated diagrams converge to diagram obtained from $Ripser$ as we increase the number of line segments ($r$). We also validate this statement by computing Bottleneck distances whose definition is given as (\cite{Kerber2016}), 
\begin{equation*}
W_{\infty}(X_{1},X_{2}) = \inf_{\eta:X_{1}\rightarrow X_{2}} \sup_{x_{1} \in X_{1}} ||x_{1}-\eta(x_{1})||_{\infty},
\end{equation*}
where $X_{1}$ and $X_{2}$ represent two different diagrams and $\eta$ is the bijections between the points of the diagrams. 
Figure~\ref{fig:distances_between_pd} shows the Bottleneck distances between the diagrams and it is seen that increasing $r$ results in smaller distances for the 8.89 cm (3.5 inch) case which means that larger values of $r$ approximates to blue one better. 
In addition, this type of behavior can be observed when we decrease the $spg$, since it will lead to larger number of groups and line segments.

%% file: section/sec-featurization.tex
\subsection{Feature Extraction Using Persistence Diagrams}
\label{sec:featurization}
In this section, we explain how to extract features from persistence diagrams using persistence landscapes, persistence images, Carlsson Coordinates, kernel for persistence diagrams and signature path of persistence path's signatures.
\subsubsection{Persistence Landscapes}
\label{sec:persistence_landscapes}
\begin{figure}
\centering
\includegraphics[width=.80\textwidth,height=.75\textheight,keepaspectratio]{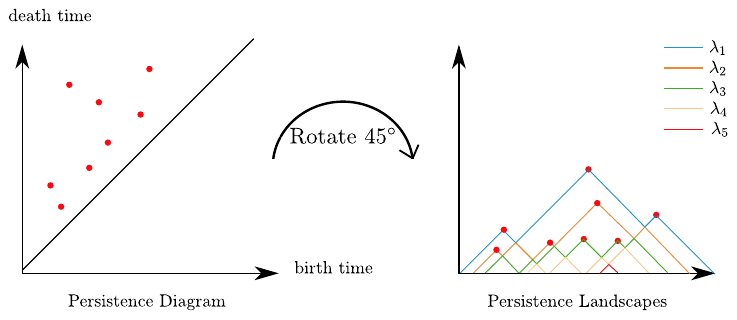}
\caption{A schematic showing the process of obtaining the landscape functions from a persistence diagram.}
\label{fig:persistence_landscapes}
\end{figure}

Persistence landscapes are functional summaries of persistence diagrams \cite{Bubenik2017}. 
They are obtained by rotating persistence diagrams by $45\degree$ clockwise, and drawing isosceles right triangles for each point in the rotated diagram \cite{Berry2018}, see Fig.~\ref{fig:persistence_landscapes} where the landscape functions are represented by $\lambda_{k}$. 
Given a persistence diagram, we define the piecewise linear functions \cite{Bubenik2017}
\begin{equation*}
  g_{(b,d)}(x) =
  \begin{cases}
    0 & \text{if $x \not\in (b,d) $} \\
  x-b & \text{if $x \in (b, \frac{b+d}{2}]$} \\
 -x+d & \text{if $x \in (\frac{b+d}{2}, d)$}
  \end{cases}
\end{equation*}
where $b$ and $d$ correspond to birth and death times, respectively. 
Figure~\ref{fig:persistence_landscapes} shows that there are several landscape functions $\lambda_k(x)$ indexed by the subscript $k \in \mathbb{N}$.  
For example, the first landscape function $\lambda_1(x)$ is obtained by connecting the topmost values of all the functions $g_{(b_{i},d_{i})}(x)$ \cite{Bubenik2017}.  
If we connect the second topmost components of $g_{(b,d)}(x)$, we obtain the second landscape function $\lambda_2$.  
The other landscape functions are obtained similarly.  
Note that the landscape functions are also piecewise linear functions.  

\textbf{Featurization of persistence landscapes:}
The persistence landscapes---$\lambda_{k}(x)$ where $x$ corresponds to the birth time---were computed using the persistence diagrams obtained from each of the embedded acceleration signals. 
Although these persistence landscapes can be utilized to featurize the persistence diagram, there is no one way to define these features. 
In this paper, we choose to extract a feature vector from the persistence landscapes by defining (a) a set of length $|\mathbb{K}|$ of the landscapes $\{\lambda_k\}_{k \in \mathbb{K}}$ where $\mathbb{K} \subset \mathbb{N}$ to work with, and (b) for the $k$th landscape, a mesh of non-empty, distinct birth times $\mathbf{b}_k = \{x_i \in \mathbb{R}\}$ where the corresponding values of the death times $\mathbf{d}_k=\{\lambda_k(x_i) \mid x_i \in \mathbf{b}_k\}$ constitute the entries of the feature vector for the $k$th landscape. 
We can then combine features from all $|\mathbb{K}|$ landscapes to obtain the full feature vector $\mathbf{d}=\{\mathbf{d}_k\}_{k \in \mathbb{K}}$ that can be used with the machine learning algorithms. 

\begin{figure}[h]
\centering
\includegraphics[width=1\textwidth,height=.75\textheight,keepaspectratio]{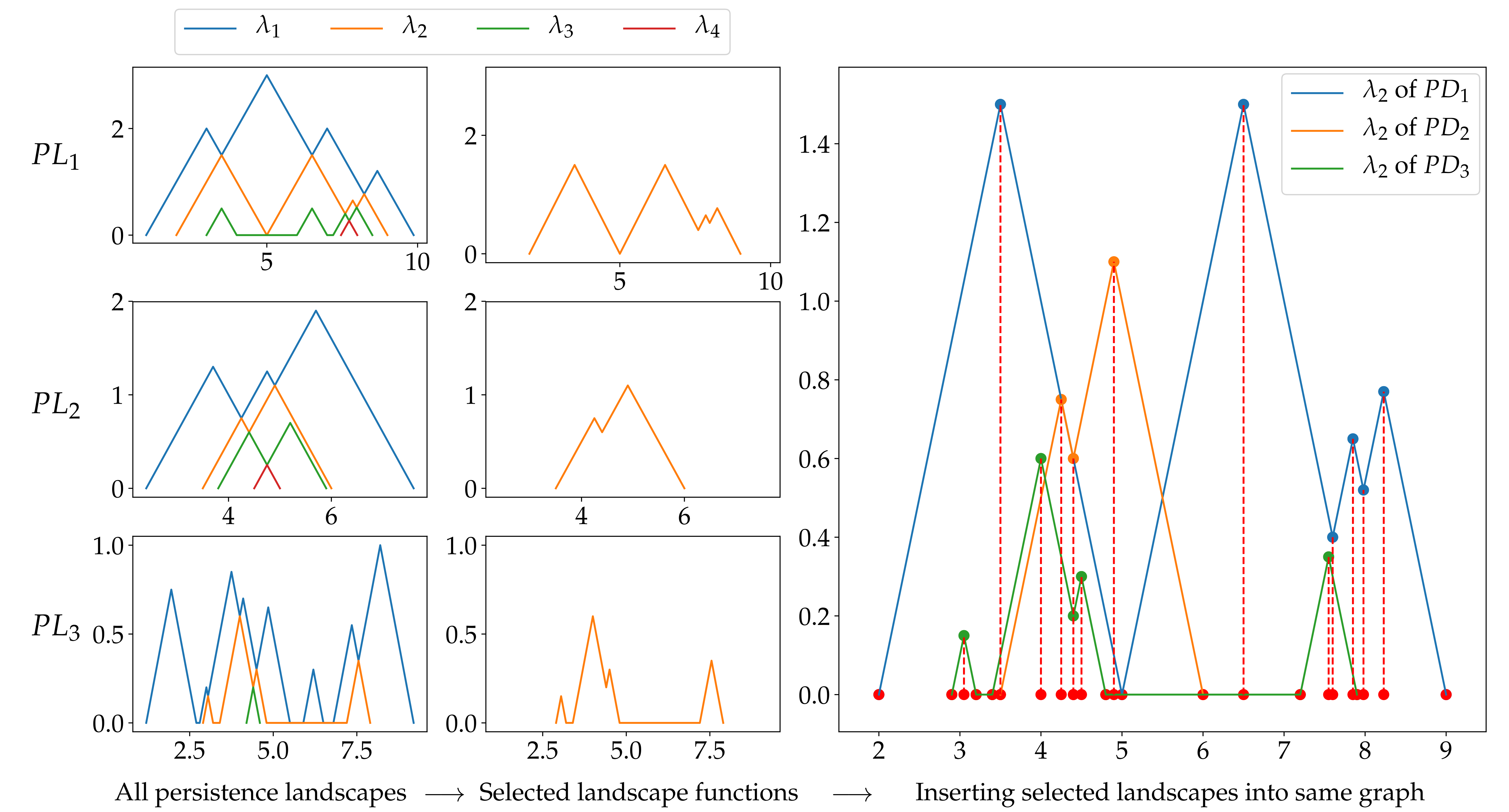}
\caption{Persistence landscape feature extraction.}
\label{fig:feature_extraction_persistence_land}
\end{figure}

Although the choice of $\mathbb{K}$, the set of landscapes to use, can be optimized using cross validation for example, in this study we set $\mathbb{K}=\{1, 2, \ldots, 5\}$ since it gives good results for our data.  
Similarly, the mesh may also be optimized in a similar way; however, this a more difficult task due to the infinite domain of $\mathbf{b}$, so we define our mesh as follows and as shown in Fig.~\ref{fig:feature_extraction_persistence_land}.

Let $\lambda_{i,j}$ be the $i$th landscape corresponding to the $j$th persistence diagram from a training set in a supervised learning setting. 
Fix $i$ and overlay the chosen landscape functions corresponding to all of the persistence diagrams in the training set.
Fig.~\ref{fig:feature_extraction_persistence_land} provides an example to this process and it selects second landscapes to extract features.
Now project all the points that define the linear pieces of each of the landscape functions onto the birth axis.
The red dots in Fig.~\ref{fig:feature_extraction_persistence_land} represents these projected points. 
Sort the projected points in ascending order and remove duplicates. 
The resulting set of points is our mesh $\mathbf{b}_i$ with length $|\mathbf{b}_i|$ 
for the $i$th landscape. 
We can repeat the same process to get the feature vector for all the $|\mathbb{K}|$ landscapes and construct the overall feature vector $\mathbf{b}$.  
We emphasize that a separate mesh is computed for each selected landscape number, and that the number of features will generally vary for each landscape function.   
Now to pull the features out of a given landscape function, we need to evaluate that function at the mesh points. 
Computationally, we efficiently accomplish that using piecewise linear interpolation functions. 

Upon extracting the features from the persistence landscapes, we construct a feature matrix which represents all the tagged feature vectors. 
For instance, Table \ref{tab:feature_matrix_PL} shows an example feature matrix obtained from the first and second landscapes corresponding to each of the $n$ persistence diagrams in the training set. 
This table shows data with two labels: $0$ for no chatter, and $1$ for chatter. 
It also denotes each feature with $y_{i,j}^{b_i}$ where $i \in \{1, 3\}$ is the landscape number, $j \in \{1, 2, \ldots, n\}$ is the corresponding persistence diagram number, while the superscript $b_i \in \{1, 2, \ldots, |\mathbf{b}_i|\}$ is the feature number corresponding to the $i$th landscape. 
These feature matrices can then be used with supervised machine learning algorithms for example to train a classifier. 

\begin{table}[h]
\caption{Feature matrix for persistence landscapes $\lambda_1$ and $\lambda_3$ corresponding to persistence diagrams $X_1$ through $X_n$. 
The entries in the cells are the values of each of the features $y_{i,j}^{b_i}$ where $i$ is the persistence landscape number, $j$ is the persistence diagram number, while $b_i$ is the feature number corresponding to the $i$th landscape.}
\label{tab:feature_matrix_PL}
\centering
\begin{tabular}{c|c|cccc|cccc}
\toprule
\multicolumn{1}{c}{Persistence Diagrams} &  \multicolumn{1}{c}{Label} &\multicolumn{4}{c}{$\lambda_{1}$} & \multicolumn{4}{c}{$\lambda_{3}$}\\
\toprule
$X_1$ &    	 1 &  	 	$y_{1,1}^{1}$ &       $y_{1,1}^{2}$&      $\dots$   &  $y_{1,1}^{|\mathbf{b}_1|}$      &       $y_{3,1}^{1}$ &     $y_{3,1}^{2}$ &     $\dots$ &  $y_{3,1}^{|\mathbf{b}_3|}$   \\

$X_2$ 	 &       0 &    	$y_{1,2}^{1}$ &       $y_{1,2}^{2}$&   		$\dots$   &  $y_{1,2}^{|\mathbf{b}_1|}$     &        $y_{3,2}^{1}$&      $y_{3,2}^{2}$&      $\dots$ &  $y_{3,2}^{|\mathbf{b}_3|}$   \\

$\vdots$ & $\vdots$ &      $\vdots$&            $\vdots$&    					 &  $\vdots$    			 &         $\vdots$  & 						  $\vdots$ &        			&   $\vdots$           \\

$X_n$ &       1 &      $y_{1,n}^{1}$ &       $y_{1,n}^{2}$&     $\dots$   &   $y_{1,n}^{|\mathbf{b}_1|}$   	 &      	$y_{3,n}^{1}$	 &		$y_{3,n}^{2}$ &    $\dots$ 	 &       $y_{3,n}^{|\mathbf{b}_3|}$		 \\   
\bottomrule
\end{tabular}
\end{table}

For our data, we computed the persistence diagrams and the corresponding first five persistence landscapes for each overhang case.  
We then randomly split the resulting landscapes into a training set ($67\%$) and test set ($33\%$), created feature matrices for each of the first five landscapes separately, and used SVM with 'rbf' kernel, Logistic Regression and Random Forest algorithms to obtain and test a classifier. 
We repeated the split-train-test process $10$ times, and every time new meshes were computed from the training sets and these same meshes were used with the corresponding test sets. 
The mean accuracy and the standard deviation of the classification computed from $10$ iterations individually using each of the first $5$ landscapes can be found in Tables~\ref{tab:result_persistence_landscape-SVM}--\ref{tab:result_persistence_landscape-RF} in the appendix. 
In the results section we utilize the results with the highest accuracy for each of the overhang cases from Tables~\ref{tab:result_persistence_landscape-SVM}--\ref{tab:result_persistence_landscape-RF} when comparing the different TDA-based featurization methods. 
\subsubsection{Persistence Images}
Persistence images are another functional summary of persistence diagrams \cite{Adams2017,Berry2018}. 
The first step in converting a persistence diagram $X=\{(b_i, d_i)\mid i \in \{1, 2, \ldots, |X| \}\}$ to persistence images is to define the linear transformation 
\begin{equation}
T(b_i,d_i)=(b_i, d_i-b_i)=(b_i, p_i),
\end{equation}
which transforms the persistence diagram from the birth-death coordinates to birth-lifetime coordinates. 
Let $D_k(x,y):\mathbb{R}^2\to \mathbb{R}$ be the normalized symmetric Gaussian centered at $(b_k, p_k)$ with standard deviation $\sigma$ according to
\begin{equation}
\label{eq:distribution_eq}
D_k(x,y)=\frac{1}{2\pi\sigma^2}e^{-\lbrack(x-b_k)^2+(y-p_k)^2\rbrack/2\sigma^2}.
\end{equation}
It was shown in \cite{Zeppelzauer2018} that the persistence images method is not very sensitive to $\sigma$, which we set to $0.1$ in this study. 
We also define a weighting function for the points in the persistence diagram $W(k)=W(b_k, p_k): (b_k, p_k)\in T(X)\to \mathbb{R}$ according to
\begin{equation}
W(k)=W(b_k, p_k) = \begin{cases}
0 & \text{ if } p_k \leq 0; \\
\frac{p_k}{b} & \text{ if } 0 < p_k < b; \\
1 & \text{ if } p_k \geq b.
\end{cases}
\end{equation}
Note that this is not the only possible weighting function, but it satisfies the requirements needed to guarantee the stability of persistence images \cite{Adams2017}: it vanishes along the horizontal axis, is continuous, and is piecewise differentiable. 
Now define the integrable persistence surface
\begin{equation}
\label{eq:persistence_surface}
S(x,y)=\sum\limits_{k\in T(X)} W(k)\,D_k(x,y).
\end{equation}
The surface $S$ can be reduced to a finite dimensional vector by defining a grid over its domain and then assigning to each box (or pixel) in this grid the integral of the surface over that pixel. 
For example, the value over the $i,j$ pixel in the grid is given by
\begin{equation}
I_{i,j}(S)=\mathop{\iint}S \, dxdy,
\end{equation}
where the integral is performed over that entire pixel. 
The persistence image corresponding to the underlying persistence diagram $X$ is the collection of all of the resulting pixels. 

\textbf{Featurization of persistence images:}
Persistence images can be used for support vector machine classification \cite{chepushtanova2015persistence}. 
The corresponding feature vector is obtained from persistence images by concatenating  the pixel values typically either row-wise or column-wise. 
The dimension of the resulting vector depends on the choice of the pixel size, i.e., the resolution of the persistence image. 
For example, let $I_{i,j}$ be a pixel in the persistence image, then a persistence image of size $100\times 100$ pixels is represented by the matrix 
\begin{equation*}
\begin{bmatrix}
I_{1,1} & I_{1,2} & \dots & I_{1,100} \\
\vdots & \ddots &  &  \\
I_{100,1} &  &  & I_{100,100}
\end{bmatrix},
\end{equation*}
and a typical feature vector is obtained by concatenating the entries of this matrix row-wise as shown by the rows in Table~\ref{table:feature_mat_PI}. 
The table shows a feature matrix where each persistence diagram is labeled either $0$ or $1$, and the corresponding feature vector is shown using entries of the form $I_{i,j}^k$, where $k \in \{1, 2, \ldots, n\}$ is the persistence diagram index while $i,j$ are row and column numbers, respectively, in the image. 
\begin{table}[h]
\caption{Feature matrix for persistence images. The notation $I_{i,j}^k$ for each feature uses $k$ to indicate the persistence diagram index, while $i,j$ are row and column numbers, respectively, in the persistence image.}
\label{table:feature_mat_PI}
\centering
\resizebox{\textwidth}{!}{\begin{tabular}{c|c|cccccccccc}
\toprule
\multicolumn{1}{c}{Persistence Diagrams} &  \multicolumn{1}{c}{Label} &\multicolumn{10}{c}{Persistence Image} \\
\toprule
$X_1$ &         1 &           $I_{1,1}^{1}$ &   $\dots$ &   $I_{1,100}^{1}$&      $I_{2,1}^{1}$   &  $\dots$  & $I_{2,100}^{1}$ &  $\dots$ & $I_{100,1}^{1}$   &  $\dots$  & $I_{100,100}^{1}$  \\

$X_2$      &       0 &        $I_{1,1}^{2}$ &   $\dots$ &   $I_{1,100}^{2}$&      $I_{2,1}^{2}$   &  $\dots$  & $I_{2,100}^{2}$ &  $\dots$ & $I_{100,1}^{2}$   &  $\dots$  & $I_{100,100}^{2}$\\

$\vdots$ & $\vdots$ &    $\ddots$      &  &                         & &&&&&   &               \\

$X_n$ &       1 &      $I_{1,1}^{p}$ &   $\dots$ &   $I_{1,100}^{p}$&      $I_{2,1}^{p}$   &  $\dots$  & $I_{2,100}^{p}$ &  $\dots$ & $I_{100,1}^{p}$   &  $\dots$  & $I_{100,100}^{p}$\\
\bottomrule
\end{tabular}}
\end{table}

We use Python's \href{https://gitlab.com/csu-tda/PersistenceImages}{PersistenceImages} package to featurize the cutting signals, and then randomly split the resulting images into $67\%$-$33\%$ train-test sets. 
The persistence images have boundaries depending on lifetime and birth time ranges. Therefore, we find the maximum lifetime and maximum birth time by checking all diagrams of a data set.
These maximum values can correspond to a point with significant lifetime and significant features can be lost if we set to boundaries to those values exactly. Accordingly, we sum each value with 1 to be able to capture all important features nearby them.
We trained a classifier using SVM and the `rbf' kernel, Logistic Regression, Random Forest and Gradient Boosting classifiers for two different pixel sizes: $0.05$ and $0.1$. 
The training and testing results for each different overhang case are available in Tables~\ref{tab:result_persistence_images-0p1}--\ref{tab:result_persistence_images-0p05} of the appendix. 
When we compare the classification accuracy for persistence images to the other featurization methods, we choose the best results from this table for each cutting configuration.

\subsubsection{Carlsson Coordinates}
Another method for featurizing persistence diagrams is Carlsson's four Coordinates \cite{Adcock2016} with the addition of the maximum persistence \cite{Khasawneh2018}, i.e, the highest off-diagonal point in the persistence diagram. 
The basic idea of Carlsson's coordinates is to utilize polynomials that (1) respect the inherent structure of the persistence diagram, and (2) that are defined on the persistence diagrams' off-diagonal points. 
Specifically, these polynomials must be able to accommodate persistence diagrams with different numbers of off-diagonal points since the persistence diagrams can vary in size even if the original datasets are of equal size. 
Further, the output of the coordinates must not depend on the order in which the off-diagonal points of a persistence diagram were stored. 
The resulting features can be computed directly from a persistence diagram $X$ according to 
\begin{equation*}
\begin{array}{rl}
f_1(X) & =  \sum{b_i(d_i-b_i)}, \\
f_2(X) & = \sum{(d_{\rm max} - d_i)(d_i - b_i)},\\
f_3(X) & = \sum{b_i^2 (d_i - b_i)^4},  \\
f_4(X) & = \sum{(d_{\rm max} - d_i)^2 (d_i-b_i)^4},\\
f_5(X) & = \max\{ (d_i - b_i) \}
\end{array}
\end{equation*}
where $d_{\rm max}$ is the maximum death time, $b_i$ and $d_i$ are, respectively, the $i$th birth and death times, and the summations and maximum are each taken over all the points in $X$.

In order to utilize Carlsson coordinates, we compute the persistence diagrams from the embedded accelerometer signals and randomly split the data into training ($67\%$) and testing ($33\%$) sets. 
We then computed all five coordinates for each diagram, and utilized SVM, Logistic Regression, Random Forest and Gradient Boosting to train a classifier. 
The feature vectors that we tested were all $\sum\limits_{i=1}^5{\tbinom{5}{i}}$ combinations of these coordinates, where the term inside the summation is $5$ choose $i$. 
This revealed which combination of features yielded the highest accuracy in each iteration.
The classification results for all of the different feature vectors are reported in Table~\ref{tab:result_carlsson_coordinates} in the appendix. 
However, in the results section we utilize the feature vectors that yielded the highest accuracy when we compare the classification results of Carlsson coordinates to the other featurization methods.

\subsubsection{Kernels for Persistence Diagrams}
In addition to featurization methods, many kernel methods have also been developed for machine learning on persistence diagrams \cite{Reininghaus2015,Kwitt2015,Zhao2019,Kusano2016, Kusano2017,Carriere2017c,Kusano2018}. 
As an example, we choose to work with the kernel introduced by~\cite{Reininghaus2015} which is defined for two persistence diagrams $X$ and $Y$ according to
\begin{equation}
\label{eq:kernel_methods}
\kappa_{\sigma}(X,Y)=\frac{1}{8\pi\sigma}\sum\limits_{z_1\epsilon X,z_2\epsilon Y} {\rm exp}\left(-\frac{||z_1-z_2||^2}{8\sigma}\right)-{\rm exp}\left(-\frac{||z_1-\hat{z}_2||^2}{8\sigma}\right),
\end{equation}
where if $z=(x,y)$, then $\hat{z}=(y,x)$, and $\sigma$ is a scale parameter for the kernel that can be used to tune the approach.  
for this study we investigated two values for this parameter: $\sigma=0.2$ and $\sigma=0.25$. 

Given either a training or a testing set $\{X_i\}_{i=1}^N$ of labeled persistence diagrams, and using Eq.~\eqref{eq:kernel_methods}, we can define the kernel matrix
\begin{equation*}
\kappa_{\sigma} =
\begin{bmatrix}
\kappa_{\sigma}(X_1,X_1) & \kappa_{\sigma}(X_1,X_2) & \dots & \kappa_{\sigma}(X_1,X_N) \\
\vdots & \ddots &  &  \\
\kappa_{\sigma}(X_N,X_1) &  &  & \kappa_{\sigma}(X_N,X_N)
\end{bmatrix}.
\end{equation*}
Note that given two persistence diagrams $X$ and $Y$ whose number of points is $|X|$ and $|Y|$, respectively, then the corresponding kernel $\kappa_{\sigma}(X, Y)$ can be computed in \textcolor{blue}{$\mathcal{O}(|X|\cdot|Y|)$} time \cite{Reininghaus2015}. 
Therefore, the computation time for kernel methods is generally high and this can complicate optimizing the tuning parameter $\sigma$. 
To emphasize the effect of the computational complexity, in this paper, the long runtime for the $5.08$ cm ($2$ inch) overhang case caused by its large number of samples has forced us to report the corresponding classification results for a smaller number of iterations than the other overhang cases and the other featurization approaches. 

For our data, we performed a $67\%$/$33\%$ train/test split of the labeled persistence diagrams. 
For each of the training and testing sets, we precomputed the corresponding kernel matrices and used Python's LibSVM \cite{LibSVM} for classification. 
For almost all but the $5.08$ cm (2 inch) overhang case where only $1$ iteration was used, we repeated the split-train-test process $10$ times and recorded the average and the standard deviation of the resulting accuracies. 
The resulting classification accuracies are reported in Table~\ref{tab:featurization_results}. 
Note that \cite{Reininghaus2015} describes another approach for training a classifier based on a measuring the distances between two kernels in combination with a $k$-Nearest Neighbor ($k$-NN) algorithm. 
However we do not explore this alternative method in this work, and only perform the computations using the kernel matrix and the LibSVM library. 

\subsubsection{Persistence Paths' Signatures}
\label{sec:persistence_paths}
Persistence paths' signatures are a recent addition to featurization tools for persistence diagrams \cite{Chevyrev2018}. 
Let $\gamma:[a, b] \to \mathbb{R}^d$ be the piecewise differentiable path given by
\begin{equation}
\gamma(t)=\gamma_{t}=[\gamma_{t}^{1},\gamma_{t}^{2},\ldots,\gamma_{t}^{d}],
\end{equation}
where each $\gamma_{t}^{i}=\gamma^{i}(t)$ is a continuous function with $t\in[a,b]$. 
The first, second, and third signatures, respectively, can be defined according to the iterated integrals \cite{Chevyrev2016}
\begin{subequations}
\begin{equation}
\label{eq:first_level}
S(\gamma)_{a,t}^{i} = \int_{a}^{t} d\gamma_{s}^{i} = \gamma_{t}^{i} - \gamma_{a}^{i},  \quad \quad  (a<s<t);
\end{equation}
\begin{equation}
\label{eq:2d_signature_path}
S(\gamma)_{a,t}^{i,j} = \int_{a}^{t} S(\gamma)_{a,s}^{i}d\gamma_{s}^{k} = \int_{a}^{t}\int_{a}^{s} d\gamma_{r}^{i}d\gamma_{s}^{j},\quad \quad (a<r<s<t);
\end{equation}
\begin{equation}
\label{eq:nd_signature_path}
S(\gamma)_{a,t}^{i,j,\ldots,k} = \int_{a}^{t} S(\gamma)_{a,s}^{i,j,\ldots,k-1}d\gamma_{s}^{k} = \int_{a}^{t} \ldots \int_{a}^{t_{2}} d\gamma_{t_{1}}^{i} \ldots d\gamma_{t_{k}}^{k}, \quad (a<t_{1}<t_{2}<\ldots<t).
\end{equation}
\end{subequations}
Other signatures are defined similarly, although, the computational cost significantly increases beyond the third level of signatures. 
The resulting path signatures can be used in classification algorithms as features. 
Looking back at persistence landscape functions in Section \ref{sec:persistence_landscapes}, we see that the $k$th landscape function $\lambda_k(t)$ can be written as a two-dimensional path
\begin{equation*}
\gamma_{t}(\lambda_k(t))=[t,\lambda_k(t)].
\end{equation*}
Therefore, we can obtain signatures from persistence landscapes and use them as features in machine learning algorithms \cite{Chevyrev2018}. 
In this paper we use up to the second level path signatures. 
Specifically, let $\lambda_{r, i}$ be the $r$th persistence landscape corresponding to the $i$th persistence diagram. 
Then the signatures that we use from the $r$th landscape function are given by $\mathbf{S}(\gamma_t(\lambda_r(t))) = [S_{r,i}^{1},S_{r,i}^{2},S_{r,i}^{1,1},S_{r,i}^{1,2},S_{r,i}^{2,1},S_{r}^{2,2}]$.
\begin{table}[h]
\caption{Feature matrix for path signatures for $n$ persistence diagrams and using the first $\lambda_1$ and second $\lambda_2$ persistence landscapes.}
\label{tab:feature_matrix_Signature}
\centering
\resizebox{\textwidth}{!}{\begin{tabular}{c|c|cccccc|cccccc}
\toprule
\multicolumn{1}{c}{Diagrams} &  \multicolumn{1}{c}{Label} &\multicolumn{6}{c}{$\lambda_{1}$} &\multicolumn{6}{c}{$\lambda_{2}$} \\
\midrule
\hline
$X_1$ & 1 &$S_{1,1}^{1}$&$S_{1,1}^{2}$ & $S_{1,1}^{1,1}$ &  $S_{1,1}^{1,2}$ & $S_{1,1}^{2,1}$ & $S_{1,1}^{2,2}$ & $S_{2,1}^{1}$ & $S_{2,1}^{2}$ & $S_{2,1}^{1,1}$ & $S_{2,1}^{1,2}$ & $S_{2,1}^{2,1}$ & $S_{2,1}^{2,2}$\\
$X_2$ 	 &0  &$S_{1,2}^{1}$&$S_{1,2}^{2}$ & $S_{1,2}^{1,1}$ &  $S_{1,2}^{1,2}$ & $S_{1,2}^{2,1}$ & $S_{1,2}^{2,2}$ & $S_{2,2}^{1}$ & $S_{2,2}^{2}$ & $S_{2,2}^{1,1}$ & $S_{2,2}^{1,2}$ & $S_{2,2}^{2,1}$ & $S_{2,2}^{2,2}$\\
$\vdots$ & $\vdots$ & $\vdots$& $\vdots$&  $\vdots$ & $\vdots$& $\vdots$ & $\vdots$ &  $\vdots$ & $\vdots$ & $\vdots$& $\vdots$ & $\vdots$  &$\vdots$ \\
$X_n$ &1 &$S_{1,n}^{1}$&$S_{1,n}^{2}$ & $S_{1,n}^{1,1}$ &  $S_{1,n}^{1,2}$ & $S_{1,n}^{2,1}$ & $S_{1,n}^{2,2}$ & $S_{2,n}^{1}$ & $S_{2,n}^{2}$ & $S_{2,n}^{1,1}$ & $S_{2,n}^{1,2}$ & $S_{2,n}^{2,1}$ & $S_{2,n}^{2,2}$\\
\hline
\end{tabular}}
\end{table}

By incorporating higher order signatures, or signatures from more landscape functions we can construct a longer feature vector for classification. 
For example, Table~\ref{tab:feature_matrix_Signature} shows the second level feature vectors computed using the first and second landscape functions for $n$ persistence diagrams.

In our experiment, we train a classifier using $75\%$ of the data, and we test using the remaining $25\%$. 
We construct a feature vector that includes up to the second path signatures for each of the first five landscape functions. 
Table~\ref{tab:result_path_signature} shows the classification accuracies for each configuration and for each landscape function. 
The best results in this table were used to compare the path signatures method to the other featurization procedures in Table~\ref{tab:featurization_results}. 

%% file: section/sec-Results.tex
\section{Results and Discussion}
\label{sec:results}
\subsection{Runtime Comparison}
\label{sec:runtime_comparison}

Runtime is a criterion for comparing the different feature extraction methods.
For the TDA-based methods, the total runtime required for classification includes ruuntime of three main steps: (1) obtaining the persistence diagrams, (2) obtain features or computing kernels, and (3) training and testing the corresponding classifier.
However, obtaining results with serial computing takes significantly larger runtime.
Therefore, we implement paralell comuting to improve the runtime. 
We used High Performance Computing Center (HPCC) of Michigan State University for parallel computing.
It includes several supercomputers which are composed of hundreds of nodes. Each node represents a computer and it has certain number of processors and RAM capacity.
Users are allowed submit multiple jobs at the same time to HPCC, and they can define number of CPU per job and memory per CPU. 
To compute the embedding parameters and persistence diagrams in parallel, we request 10 CPU per job and 2GB of memory per CPU. 
The number of jobs submitted to HPCC changes depending on the number of time series of the overhang distance.

Embedded time series are subsampled such that every $10^{th}$ point is taken into account to compute persistence diagrams.
We recorded the times to complete persistence diagram computation of all overhang cases and report them in Tab.~\ref{tab:comparison_paralell_serial}. 
It also includes times for serial computing where one persistence diagram is computed at a time. It is seen that parallel computing reduces the computation time significantly, although most part of the runtimes for parallel computing is the queue time.
Parallel computing can also be performed with some workstations available in market without having need of expensive supercomputers that HPCC have. Entry level workstations with CPU having 64 cores and 512 GB of RAM can be afforded by small workshops. 
 
\begin{table}[h]
\centering
\caption{Comparison of runtimes (seconds) for embedding parameters and persistence diagram computation of all overhang cases with parallel and serial computing.}
\label{tab:comparison_paralell_serial}
\resizebox{0.75\textwidth}{!} { 
\begin{tabular}{c|cc|cc|cc|cc}
\multicolumn{1}{c}{}&\multicolumn{2}{c}{\makecell{5.08 cm\\ (2 inch)}} & \multicolumn{2}{c}{\makecell{6.35 cm\\(2.5 inch )}} & \multicolumn{2}{c}{\makecell{8.89 cm\\ (3.5 inch)}}  &\multicolumn{2}{c}{\makecell{11.43 cm\\ (4.5 inch)}} \\
\hline
\multicolumn{1}{c|}{} & Parallel & Serial & Parallel & Serial & Parallel & Serial & Parallel & Serial \\
\hline
\makecell{Persistence \\Diagram} & 9420 & 84346 & 3448 & 23570 & 2073 & 11319 & 4819 & 37617\\
\hline
\end{tabular}}
\end{table}

Despite long computation time for persistence-based methods, we note that after obtaining the persistence diagrams, they can be saved and used in multiple TDA-based classification methods. 
In addition, it was observed that delay and embedding dimension parameters do not change significantly for changing time series of a same cutting configuration. Parameters for embedding can be computed in the training phase of a classifier and they will be used in the test phase. 
Therefore, once these diagrams are computed, the time required for featurization and classification would be a fraction of the the ones reported in Tab.~\ref{tab:comparison_paralell_serial}. 
Another point we wish to emphasize is that the most computationally expensive step is that of training a classifier. 
Once a classifier is trained, which can be done offline, the effort in classifying incoming streams of data is much smaller because a much smaller set of persistence diagrams and features are needed.
Therefore, we compare the run times is needed for a single time series with different methods.
Table~\ref{tab:single_ts_time_comparison} provide runtimes for embedding parameters computations and  persistence diagram computation with different methods.

\begin{table}[h]
\centering
\caption{Runtime (seconds) for embedding parameter computation and persistence diagram computation of a single time series with different methods.}
\label{tab:single_ts_time_comparison}
\resizebox{\textwidth}{!} { 
\begin{tabular}{c|c|ccccccc}
\multicolumn{1}{c|}{} & \multicolumn{1}{c|}{\makecell{Embedding \\ Parameters}}& \multicolumn{7}{c}{\makecell{Persistence Diagram}} \\
\hline
\multicolumn{1}{c|}{} & \multicolumn{1}{c|}{}& \multicolumn{1}{c|}{\makecell{Number of \\points$\approx1000$}} & \multicolumn{3}{c|}{\makecell{Number of points$\approx100$}}& \multicolumn{3}{c}{\makecell{Number of points$\approx300$}}\\
\hline
\multicolumn{1}{c|}{\makecell{Overhang \\ Distances}} & \multicolumn{1}{c|}{\makecell{Delay and \\Dimension}}& \multicolumn{1}{c|}{\makecell{Subsampled \\Point Cloud}} & \makecell{Greedy \\ Permutation \\ $n_{perm} = 100$} & \makecell{B\'ezier \\$r=1$\\ $spg=100$ \\ (Serial)} & \multicolumn{1}{c|}{\makecell{B\'ezier \\$r=1$\\ $spg=100$ \\ (Parallel)}} & \makecell{Greedy \\ Permutation \\ $n_{perm} = 300$} & \makecell{B\'ezier \\$r=3$\\ $spg=100$ \\ (Serial)} &  \makecell{B\'ezier \\$r=3$\\ $spg=100$ \\ (Parallel)} \\
\hline
\makecell{5.08 cm  \\ (2 inch)}   & 242.63  &\multicolumn{1}{c|}{266.95} & 0.2  & 106.00 &\multicolumn{1}{c|}{0.93} &   3.62 & 569.78  & 6.21\\
\makecell{6.35 cm  \\(2.5 inch)}  & 191.10  &\multicolumn{1}{c|}{208.95} & 0.29 & 106.15 &\multicolumn{1}{c|}{0.85} &   3.79 & 538.98 &  6.19 \\
\makecell{8.89 cm  \\ (3.5 inch)} & 166.79  &\multicolumn{1}{c|}{296.21} & 0.18 & 96.00  &\multicolumn{1}{c|}{0.72} &    3.87 & 541.62 &  6.48 \\
\makecell{11.43 cm \\ (4.5 inch)} & 200.51  &\multicolumn{1}{c|}{276.38} & 0.17 & 113.86 &\multicolumn{1}{c|}{0.77} &  3.53  &600.29 &  6.99 \\
\hline
\end{tabular}}
\end{table}

First column in Tab.\ref{tab:single_ts_time_comparison} represent the runtimes of computing embedding dimension and delay parameter. When these parameters computed, they can be saved and used in embedding time series
In the second column of Tab.~\ref{tab:single_ts_time_comparison}, the runtime of persistence diagram computation of subsampled point cloud is given. In this way, we use nearly 1000 points from the embedded time series to compute persistence diagrams.  
However, computation time for a persistence diagrams of a point cloud with that size is still high as seen from Tab~\ref{tab:single_ts_time_comparison}. 
Therefore, we also employ greedy permutation subsampling and B\'ezier curve approximation technique.
Greedy permutation option of $Ripser$ package is utilized and it is a method that subsamples the point cloud and computes the persistence diagrams with less number of points. 
$n_{perm}$ is a parameter that defines the number of points selected by greedy permutation algorithm. 
We chose $100$ and $300$ points for this option and reported runtimes for resulting persistence diagrams.
In Tab.~\ref{tab:single_ts_time_comparison}, we grouped the runtimes with respect to the number of points used in the corresponding method.
It is seen that B\'ezier curve approximation with $r=1$ and $r=3$ uses approximately $100$ and $300$ points as in the case of greedy permutation.  
We provide runtime for both serial and parallel computing in Tab.~\ref{tab:single_ts_time_comparison}.
Parallel computing can only be applied to B\'ezier curve approximation among the methods of persistence diagram computation given in Tab.\ref{tab:single_ts_time_comparison}.
The reason is that persistence diagrams are obtained directly from the $Ripser$ package for other methods.
However, the steps of B\'ezier curve approximation method, computation of coefficients for the line segments and the distance matrix between these line segments can be performed in parallel.
In these two steps, a job can only compute coefficients of the lines in a single group or a distance between two lines. 
The number of the jobs will be equal to group number and number of combinations between lines for coefficient computation and distance matrix computation, respectively.
Ideally, all jobs for a step can be computed simultaneously, if there is no queue time.
Therefore, we recorded runtimes for computation of coefficients of the line segments in a single group, computation of a distance between two line segments and persistence diagram from a distance matrix individually.
Then, we sum them up and reported in Tab.~\ref{tab:single_ts_time_comparison}.
Combining parallel computing with B\'ezier curve approximation reduces the runtime significantly. 
Moreover, it is seen that the fastest method is the greedy permutation with $n_{perm}=100$ and B\'ezier curve approximation computed in parallel places second.
Both method is able to complete the diagram computation in less than a second, while runtime gets larger with increasing $n_{perm}$ and $r$ parameters.

Table~\ref{tab:single_ts_classification_time_comparison} provides the times required to complete classification of a single time series. 
To be fair in comparison between runtime of WPT/EEMD and TDA-based methods, we assumed that the classifier is already trained and required parameters for all methods are selected. 
It is seen that WPT is the fastest method and EEMD places second. 
Runtime for TDA-based method is comparable to the ones for EEMD.
Further the WPT and EEMD methods use codes that have been highly optimized, whereas the TDA-based methods are still under active research with huge future potential for improved optimization. 
We believe that the runtimes for TDA-based method can be further decreased with optimization.

\begin{table}[h]
\centering
\caption{Runtime (seconds) for performing classification with a single time series for TDA based methods and signal decomposition based ones.}
\label{tab:single_ts_classification_time_comparison}
\resizebox{0.8\textwidth}{!} { 
\begin{tabular}{c|cccc|cc}
\multicolumn{1}{c|}{} & \multicolumn{4}{c|}{\makecell{Topological Data Analysis}}& \multicolumn{2}{c}{Signal Decomposition} \\
\hline
\multicolumn{1}{c|}{\makecell{Overhang \\ Distances}} & \makecell{Persistence \\Landscapes}& \makecell{Template \\Functions} & \makecell{Carlsson \\ Coordinates} & \multicolumn{1}{c|}{\makecell{Persistence \\Images}} & WPT & EEMD  \\
\hline
\makecell{5.08 cm  \\ (2 inch)}   & 1.01  &0.97 & 0.97 & 0.97 &0.03 &   0.52 \\
\makecell{6.35 cm  \\(2.5 inch)}  & 0.92  &0.90 & 0.90 & 0.87 &0.08 &   0.65  \\
\makecell{8.89 cm  \\ (3.5 inch)} & 0.81  &0.81 & 0.76 & 0.76  &0.09 &  0.70  \\
\makecell{11.43 cm \\ (4.5 inch)} & 0.87  &0.81 & 0.81 & 0.81 &0.06 &  0.52   \\
\hline
\end{tabular}}
\end{table}

\subsection{Classification Scores}
\label{sec:cl_score}
This section presents the classification accuracies for all the methods that are introduced in Section \ref{sec:method} and compares them to the results in Ref.~\cite{Yesilli2020} which uses the Wavelet Packet Transform (WPT) and the Ensemble Empirical Mode Decomposition (EEMD). 
The latter two methods are used for comparison since they are some of the currently most prominent methods for chatter identification using supervised learning.
For persistence images, Template Functions and Carlsson Coordinates, we applied four different classifier and these are Support Vector Machine (SVM), Logistic Regression (LR), Random Forest (RF) and Gradient Boosting (GB) algorithms. All of these classifier except Gradient Boosting is used in Persistence Landscape method, while Kernel method and Persistence Paths results are obtained with LibSVM and SVM classifiers, respectively.
The classification results are summarized in Table~\ref{tab:featurization_results} where for each cutting configuration the best results of the classification algorithms for each method are included.
Table~\ref{tab:featurization_results} also includes the classification results obtained using a new TDA approach, which is not included in Section \ref{sec:method}, based on template functions \cite{Perea2019}. 
In this table, the best accuracy for each dataset is highlighted in green. 
Further, methods whose accuracy are within one standard deviation of the best result in the same category are highlighted in blue. 

\begin{table}[h]
\centering
\begin{threeparttable}
\caption{Comparison of results for each method where WPT is the Wavelet Packet Transform, and EEMD stand for Ensemble Empirical Mode Decomposition. }
\label{tab:featurization_results}
\resizebox{\textwidth}{!}{\begin{minipage}{\textwidth}
\begin{tabular}{|c|c|c|c|c|c|c|c|c|c|c|}
\hline
\makecell{Overhang\\Length\\cm\\(inch)} & \makecell{Persistence\\Landscapes} & \makecell{Persistence\\Images} & \makecell{Template\\Functions} & \makecell{Carlsson\\Coordinates} & \makecell{Kernel\\Method} & \makecell{Persistence\\Paths} & WPT & EEMD\\
\hline
\makecell{5.08\\(2)}   &\cellcolor[RGB]{75,228,141}$\SI{96.8}{\percent}$ & \cellcolor[RGB]{204,229,255}$\SI{96.4}{\percent}$&$\SI{91.5}{\percent}$ & $\SI{93.6}{\percent}$ & $\SI{74.5}{\percent}$* & $\SI{83.0}{\percent}$ & $\SI{93.9}{\percent}$ & $\SI{84.2}{\percent}$\\
\hline
\makecell{6.35\\(2.5)}	&$\SI{88.6}{\percent}$ & $\SI{85.8}{\percent}$&$\SI{89.3}{\percent}$ & $\SI{86.3}{\percent}$ & $\SI{58.9}{\percent}$ & $\SI{84.2}{\percent}$ & \cellcolor[RGB]{75,228,141}$\SI{100.0}{\percent}$ & $\SI{78.6}{\percent}$ \\
\hline
\makecell{8.89\\(3.5)}	&\cellcolor[RGB]{204,229,255}$\SI{92.2}{\percent}$ & \cellcolor[RGB]{204,229,255}$\SI{93.0}{\percent}$&$\SI{83.9}{\percent}$ & \cellcolor[RGB]{75,228,141}$\SI{95.7}{\percent}$ & $\SI{87.0}{\percent}$ & $\SI{85.9}{\percent}$ & $\SI{84.0}{\percent}$ & $\SI{90.7}{\percent}$ \\
\hline
\makecell{11.43\\(4.5)}	&$\SI{68.6}{\percent}$ & $\SI{72.5}{\percent}$&$\SI{65.1}{\percent}$ & $\SI{72.2}{\percent}$ & $\SI{59.3}{\percent}$ & $\SI{70.0}{\percent}$ & \cellcolor[RGB]{75,228,141}$\SI{87.5}{\percent}$ & \cellcolor[RGB]{204,229,255}$\SI{79.1}{\percent}$\\
\hline
\end{tabular}
\end{minipage}}
\begin{tablenotes}
\small
\item *This result belongs to only the first iteration for the $5.08$ cm (2 inch) overhang case. 
\end{tablenotes}
\end{threeparttable}
\end{table}
Table~\ref{tab:featurization_results} shows that the WPT approach yields the highest classification accuracy for the $6.35$ cm ($2.5$ inch) and the $11.43$ cm ($4.5$ inch) overhang cases. 
However, we also see that for the $5.08$ cm ($2$ inch) and the 8.89 cm (3.5 inch) case, persistence landscape and Carlsson Coordinates yield the highest accuracy, respectively. 
For the $6.35$ cm ($2.5$ inch) overhang case we point out that the number of time series is small. 
Specifically, for this case, less than $10$ time series were divided into small pieces and used as the test set, see Table~\ref{tab:chatter_case_number}. 
Therefore, the $100\%$ classification accuracy using WPT for this case does not represent a robust result. 
Nevertheless, for the same case Table~\ref{tab:featurization_results} shows that the TDA methods based on persistence landscapes, persistence images, template functions, Carlsson coordinates, and persistence paths yield better results than EEMD---a leading approach for chatter detection. 
For the $8.89$ cm ($3.5$ inch) case, Carlsson coordinates method yields the highest mean accuracy of $95.7\%$, placing ahead of both WPT and EEMD. 
Further, the other TDA-based method for this cutting configuration score classification accuracies of at least $83.9\%$. 
In addition, WPT and EEMD includes manual preprocessing which requires extra time and skill \cite{Yesilli2020}.
Therefore, the automation of these processes are not feasible, while all steps in TDA-based feature extraction can be fully automatized.

\begin{figure}[h]
\centering
\includegraphics[width=1\textwidth]{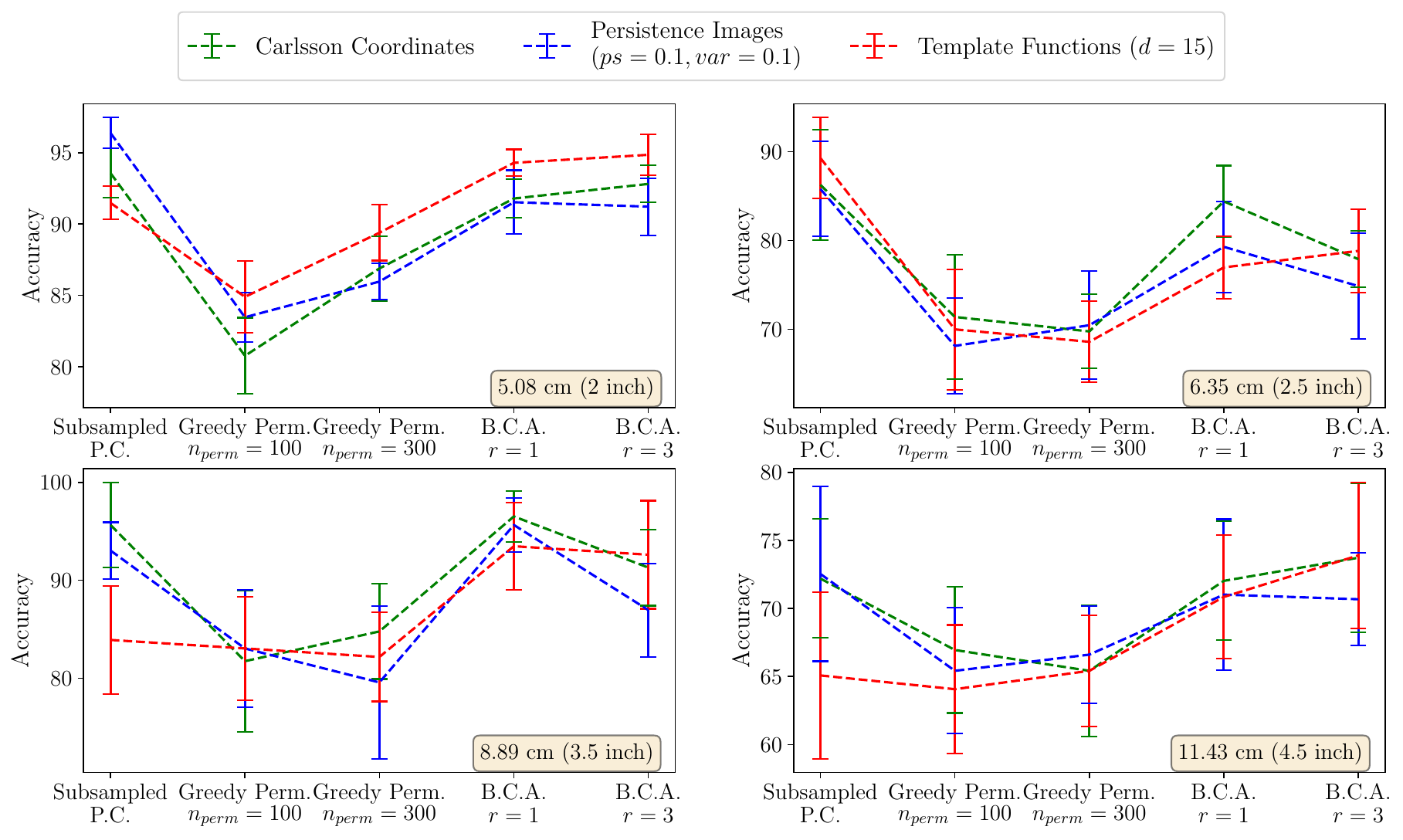}
\caption{Classification performance of persistence diagrams obtained with different methods for all overhang cases.}
\label{fig:Score_comparison}
\end{figure}

We also compare the performance of different persistence diagram computation methods whose runtime is compared in Sec.\ref{sec:runtime_comparison}. Figure~\ref{fig:Score_comparison} shows you the mean classification accuracies and error for the persistence diagrams obtained with the ways shown in Tab.~\ref{tab:single_ts_time_comparison}. Subsampling point cloud at every $10^{th}$ point is the first method to compute persistence diagrams.  
Tab~\ref{tab:single_ts_time_comparison} shows that B\'ezier curve approximation has slightly larger computation time compared to greedy permutation subsampling method when it is computed in parallel.
However, it is seen that for all overhang cases B\'ezier curve approximation method results in higher accuracy compared to greedy permutation.
Also, its result are the closest results to the ones obtained from subsampled point cloud.

Increasing number of points and greedy permutation or increasing number of line segments ($r$) generated for a group does not always yield higher accuracy as we see from Fig.~\ref{fig:Score_comparison}. 
The reason could be that increasing the number of points or the number of line segments can cause more topological noise on persistence diagrams. 
For example, Fig.~\ref{fig:Comparison_of_diagrams} shows that the new point appears closed to significant feature, which is the point with highest lifetime, as we increase $r$ from $5$ to $7$. Therefore, this could cause small drops in accuracy for some overhang cases as seen from the Fig.~\ref{fig:Score_comparison}.

Greedy permutation and B\'ezier curve approximation method can provide persistence diagrams in less than 1 second without optimization. 
For greedy permutation method, we can not apply any optimization, however coefficient computation for the line segments in B\'ezier curve approximation method can be optimized.
This could further decrease the runtimes and opens the possibility for exploring in-situ chatter detection using TDA-based methods especially with properly optimized algorithms.  

%% file: section/sec-Conclusion.tex
\section{Conclusion}
\label{sec:Conclusion}
Two of the most common methods for chatter detection are based on extracting features from the time series using the WPT and the EEMD methods. 
However, even after obtaining a tagged time series, utilizing either of these methods requires the analysis of at least a subset of the available time series to identify which parts of the signal's decomposition are the most informative when training a classifier. 
In addition to requiring highly trained individuals to perform this critical step, both WPT and EEMD are then locked into a small set of informative packets or IMFs which limits the transfer learning ability of the resulting classifiers \cite{Yesilli2020}. 

In contrast to the WPT and EEMD methods, we use persistent homology---the most prominent analysis tool from TDA---to obtain a summary of the persistent topological features of the data. 
These are based on the global structure of the point cloud embedding of the acceleration signals in a turning experiment; 
therefore, upon obtaining a persistence diagram, there is no manual work involved in selecting the features from the persistence diagram.  
Since working directly with the resulting persistence diagrams is difficult, we investigated using the leading tools for feature extraction from persistence diagrams. 
The featurization methods that we studied are based on persistence landscapes, persistence images, Carlsson Coordiantes, a kernel method, template functions, and persistence paths' signatures. 
The resulting features are then combined with several classification algorithms for training a classifier. 
The classification results are then computed from multiple split-train-test sets, and the resulting mean accuracies as well as the corresponding standard deviation are recorded for each method and for each cutting configuration. 

Tables~\ref{tab:featurization_results} summarizes the classification accuracy of all the TDA-based tools as well as their WPT and EEMD counterparts. 
In terms of the classification accuracy across the different cutting configurations we note the Carlsson coordinates and persistence landscapes approaches yield the best accuracies for two cases, and the former has the smallest runtime in comparison to the other TDA tools. On the other hand, in remaining cases, WPT yielded the best accuracies.
Table \ref{tab:featurization_results} shows that WPT yields an accuracy of $100\%$ (with a standard deviation of zero) for the $6.35$ cm ($2.5$ inch) overhang case; however, as pointed out in Section \ref{sec:results} and Table~\ref{tab:chatter_case_number}, the size of the test set for this cutting configuration is too small which casts some doubts about the robustness of this result. 
Nevertheless, for this case both template functions and Carlsson coordinates still yield at least $86\%$ classification accuracy. 
Specifically, Table~\ref{tab:featurization_results} shows that for the $6.35$ cm (2.5 inch) case, persistence landscapes, and Carlsson coordinates yield accuracies that are off by $13.7\%$, and $10.7\%$, respectively, from the WPT result. 
Similarly, for the $11.43$ cm ($4.5$ inch) case, Carlsson Coordinates and persistence images are within $13.3\%$ and $13\%$ of the EEMD result. 
For the 5.08 cm (2 inch) and $8.89$ cm ($3.5$ inch) overhang case, persistence landscapes and Carlsson coordinates yield the highest accuracy scoring $96.8\%$ and $95.7\%$, respectively, with tight error bounds that do not enclose the WPT accuracies. 

As for runtime comparisons,  we see dramatic decrease in persistence diagram computation when we computed persistence diagram in parallel as seen from Tab.~\ref{tab:comparison_paralell_serial}. 
Parallel computing reduces the runtime for computing persistence diagram at least $82\%$. 
Table~\ref{tab:single_ts_classification_time_comparison} shows that WPT is the fastest followed by EEMD; 
however, the reported time for TDA-based approach is comparable to the ones of EEMD.
We can reduce the runtime for computation of persistence diagram of a single time series to less than 1 seconds using Greedy permutation and B\'ezier curve approximation method with combination of parallel computing. Fig.\ref{fig:Score_comparison} also shows that B\'ezier curve approximation method results in larger accuracies compared to Greedy permutation. Therefore, combination of parallel computing and B\'ezier curve approximation will make TDA-based approaches feasible option for online chatter detection.

To summarize, we show that persistence features are appropriate for chatter detection in cutting processes. 
These features have the potential to lower the barrier to entry when tagging cutting signals as chatter or chatter free because no manual pre-processing is needed before extracting and using the features in the persistence diagram. 
We also note that after obtaining a classifier, the time required for the classification of new incoming data will be greatly reduced, thus opening the door for future implementation of TDA methods in-situ for chatter detection and mitigation. We also believe that applying TDA based feature extraction techniques to different machining processes and machine tools can be included in future work.

%% file: section/sec-appendix.tex
\appendixpage
\appendix

\section{Simplicial complexes}
\label{sec:SimplicialComplexes}
\begin{figure}[h!]
\centering
\includegraphics[width=0.85\textwidth,height=.75\textheight,keepaspectratio]{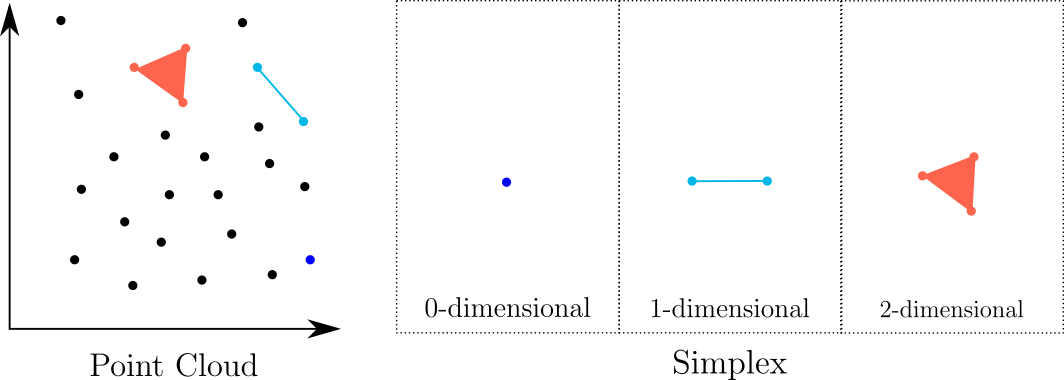}
\caption{Formation of simplicial complexes from point cloud.}
\label{fig:simplicial_complex}
\end{figure}

Let $\{u_{0},\ldots,u_{k}\} \in \mathbb{R}^{d}$ be a  set of data points, and the vectors defined between these data points ($u_{1}$-$u_{0}$,$u_{2}$-$u_{0}$,\ldots,$u_{k}$-$u_{0}$) are linearly independent.
A geometric \textit{k-}simplex, $\sigma$ 
is a set of all points in $\mathbb{R}^{d}$ such that
$\sum_{j=0}^{k} \lambda_{j}u_{j}$ where $\sum_{j=0}^{n} \lambda_{j} =1$ and $\lambda_{j}\geq 0$ for all j. 
Figure~\ref{fig:simplicial_complex} provides illustrations for $0,1,$ and $2-$simplex. 
Each data point on a point cloud is represented as 0-dimensional simplex and they are called vertices. 
When two vertices are connected, an edge is formed and it is 1-dimensional simplex. 
Connection of three vertices will form $2-$simplex which is a triangle.
Simplicies spanned by any subset of $u_{0},\ldots,u_{k}$ are the faces of $\sigma$.
In general, $n-$simplex contains $n+1$ vertices, and the set of these simplices, are called geometric simplicial complexes, $K$, if the following two conditions are satisfied \cite{Munkres2018}: 
1) If $\sigma \in K$, then faces of $\sigma$ are also in $K$, 
2) If two $n-$simplex, $\sigma_{1}$ and $\sigma_{2}$ are in $K$, then the intersection of them is either common face or empty.
The dimension of the simplicial complex, $K$ is equal to the largest dimension of its simplices. 

\section{Expressions for persistence paths' signatures}
\label{appndx:path_signature_expressions}
Let the $k$th landscape functions be $\lambda_{k}(x)=y$ where $x$ and $y$ represent coordinates along the birth time and the persistence axes. 
Since the persistence landscapes are piecewise linear functions, we can write them in closed form in terms of the nodes $\{x_i, y_i\}_{i=1}^n$ that define the boundaries of each of their linear pieces according to
\begin{equation}
\label{eq:intermsof_x_y}
\lambda_k(t) = \begin{cases}
\frac{y_{i+1}-y_{i}}{x_{i+1}-x_{i}} t + \frac{y_{i+1}(x_{i+1}-x_{i})+x_{i+1}(y_{i}-y_{i+1})}{x_{i+1}-x_{i}}, &
\text{ for }i \in \{1,2,\ldots, n\} \text{ and } t \in [x_1, x_n], \\
0 , &\text{ otherwise }.
\end{cases}
\end{equation} 
The corresponding path is given by
\begin{equation}
P=[P_{t}^{1},P_{t}^{2}] = \left[t,\frac{y_{i+1}-y_{i}}{x_{i+1}-x_{i}} t + \frac{y_{i+1}(x_{i+1}-x_{i})+x_{i+1}(y_{i}-y_{i+1})}{x_{i+1}-x_{i}}\right],
\end{equation}
and its differential is
\begin{equation}
\label{eq:derivative}
dP = [dP_{t}^1,dP_{t}^2] = [dt, \frac{y_{i+1}-y_{i}}{x_{i+1}-x_{i}} dt].
\end{equation}
Using the above definitions, we can derive general expressions for the first and second level signatures, respectively, according to
\begin{subequations}
\begin{equation}
\label{eq:firstlevel_firstpath}
S_{1} =  \int_{x_{1}}^{x_{n}} dP_{t}^1 = \int_{x_{1}}^{x_{n}} dt = x_{n} - x_{1}
\end{equation}
\begin{equation}
\label{eq:firstlevel_secondpath}
S_{2} =  \int_{x_{1}}^{x_{n}} dP_{t}^2 = \int_{x_{1}}^{x_{2}} \frac{y_{2}-y_{1}}{x_{2}-x_{1}} dt +  \ldots + \int_{x_{i}}^{x_{i+1}} \frac{y_{i+1}-y_{i}}{x_{i+1}-x_{i}} dt + \ldots +\int_{x_{n-1}}^{x_{n}} \frac{y_{n}-y_{n-1}}{x_{n}-x_{n-1}} 
\end{equation}
\end{subequations}
The equations for the second level signatures is provided in Table~\ref{tab:signature_formulas}.

\begin{table} 
\centering
\caption{Closed-form expressions for the first and second level signature paths for landscape functions.}
\label{tab:signature_formulas}
\resizebox{\textwidth}{!}{\begin{tabular}{c  l}
\hline
\addlinespace[0.2cm]
\multicolumn{1}{c}{Signature} &  \multicolumn{1}{c}{Equation} \\
\addlinespace[0.2cm]
\hline
\addlinespace[0.5cm]
$S_{1}$ & $\begin{aligned}\int_{x_{1}}^{x_{n}} dP_{t}^1 = \int_{x_{1}}^{x_{n}} dt = x_{n} - x_{1}\end{aligned}$\\
\addlinespace[0.5cm]
$S_{2}$ & $\begin{aligned}\int_{x_{1}}^{x_{n}} dP_{t}^2 = \int_{x_{1}}^{x_{2}} \frac{y_{2}-y_{1}}{x_{2}-x_{1}} dt + \ldots + \int_{x_{i}}^{x_{i+1}} \frac{y_{i+1}-y_{i}}{x_{i+1}-x_{i}} dt + \ldots + \int_{x_{n-1}}^{x_{n}} \frac{y_{n}-y_{n-1}}{x_{n}-x_{n-1}}\end{aligned}$ \\ 
\addlinespace[0.5cm]
$S_{1,1}$ & $\begin{aligned}\int\int{dP_{t}^1}dP_{t}^1 = \int_{x_{1}}^{x_{n}}\bigg(\int_{x_{1}}^{t_{2}} {dt_{1}}\bigg) dt_{2} = \int_{x_{1}}^{x_{n}} (t_{2}-x_{1}) dt_{2} = \bigg(\frac{t_{2}^{2}}{2}-x_{1}t_{2}\bigg)\bigg|_{t_{2}=x_{1}}^{t_{2}=x_{n}}\end{aligned}$\\
\addlinespace[0.5cm]
$S_{1,2}$ & $\begin{aligned}\int\int{dP_{t}^1}dP_{t}^2 &= \int_{x_{1}}^{x_{n}} \bigg(\int_{x_{1}}^{t_{2}} {dt_{1}}\bigg) dP_{t}^2 dt_{2} \\ &= \int_{x_{1}}^{x_{2}} (t_{2}-x_{1}) \frac{y_{2}-y_{1}}{x_{2}-x_{1}} dt_{2}+ \ldots +\int_{x_{i}}^{x_{i+1}} (t_{2}-x_{1}) \frac{y_{i+1}-y_{i}}{x_{i+1}-x_{i}} dt_{2}+ \ldots + \int_{x_{n-1}}^{x_{n}} (t_{2}-x_{1}) \frac{y_{n}-y_{n-1}}{x_{n}-x_{n-1}}dt_{2}\end{aligned}$ \\
\addlinespace[0.5cm]
$S_{2,1}$ &
$\begin{aligned}
\int\int{dP_{t}^2}dP_{t}^1 & =  \int_{x_{1}}^{x_{n}} \bigg(\int_{x_{1}}^{t_{2}} \frac{y_{i+1}-y_{i}}{x_{i+1}-x_{i}} {dt_{1}}\bigg) dt_{2}  = \int_{x_{1}}^{x_{n}} \frac{y_{i+1}-y_{i}}{x_{i+1}-x_{i}} \bigg(\int_{x_{1}}^{t_{2}} {dt_{1}}\bigg) dt_{2} \\
& = \int_{x_{1}}^{x_{2}} \frac{y_{2}-y_{1}}{x_{2}-x_{1}} \bigg(\int_{x_{1}}^{t_{2}} {dt_{1}}\bigg) dt_{2} + \ldots + \int_{x_{i}}^{x_{i+1}} \frac{y_{i+1}-y_{i}}{x_{i+1}-x_{i}} \bigg(\int_{x_{i}}^{t_{2}} {dt_{1}}\bigg) dt_{2} +\ldots+ \int_{x_{n-1}}^{x_{n}} \frac{y_{n}-y_{n-1}}{x_{n}-x_{n-1}} \bigg(\int_{x_{n}}^{t_{2}} {dt_{1}}\bigg) dt_{2}
\end{aligned}$
\\
\addlinespace[0.5cm]
$S_{2,2}$ & $\begin{aligned}
\int\int{dP_{t}^2}dP_{t}^2 & = \int_{x_{1}}^{x_{n}} \bigg(\int_{x_{1}}^{t_{2}} \frac{y_{i+1}-y_{i}}{x_{i+1}-x_{i}} {dt_{1}}\bigg) \frac{y_{i+1}-y_{i}}{x_{i+1}-x_{i}} dt_{2}  = \int_{x_{1}}^{x_{n}} \bigg(\frac{y_{i+1}-y_{i}}{x_{i+1}-x_{i}}\bigg)^{2} \bigg(\int_{x_{1}}^{t_{2}} {dt_{1}}\bigg) dt_{2} \\
& = \int_{x_{1}}^{x_{2}} \bigg(\frac{y_{2}-y_{1}}{x_{2}-x_{1}}\bigg)^{2} \bigg(\int_{x_{1}}^{t_{2}} {dt_{1}}\bigg) dt_{2} + \ldots + \int_{x_{i}}^{x_{i+1}} \bigg(\frac{y_{i+1}-y_{i}}{x_{i+1}-x_{i}}\bigg)^{2} \bigg(\int_{x_{i}}^{t_{2}} {dt_{1}}\bigg) dt_{2} +\ldots+ \int_{x_{n-1}}^{x_{n}} \bigg(\frac{y_{n}-y_{n-1}}{x_{n}-x_{n-1}}\bigg)^{2} \bigg(\int_{x_{n}}^{t_{2}} {dt_{1}}\bigg) dt_{2}
\end{aligned}$
\end{tabular}}
\end{table}

\section{Classification results}
\label{appndx:classification_results}

\begin{table}
\centering
\caption{Persistence Landscape Results with Support Vector Machine (SVM) classifier. Landscape number column represents which landscapes was used to extract features.}
\label{tab:result_persistence_landscape-SVM}
\begin{tabular}{|c|c|c|c|c|}
\hline
\multicolumn{1}{|c|}{} & \multicolumn{2}{c|}{2 inch} & \multicolumn{2}{c|}{2.5 inch} \\
\hline
\multicolumn{1}{|c|}{\makecell{Landscape \\ Number}} & Test Set &Training Set & Test Set& Training Set \\
\hline
1 &  \cellcolor[RGB]{75,228,141}$\SI{96.8  \pm  1.5}{\percent}$   & $\SI{96.7  \pm  0.5}{\percent}$   & $\cellcolor[RGB]{204,229,255}\SI{87.0  \pm  4.8}{\percent}$   & $\SI{86.9  \pm 2.4}{\percent}$ \\
2&   $\SI{86.6  \pm  1.7}{\percent}$   & $\SI{91.5  \pm  0.8}{\percent}$   & \cellcolor[RGB]{204,229,255}$\SI{87.0  \pm  3.8}{\percent}$   & $\SI{86.0  \pm 2.0}{\percent}$ \\
3&   $\SI{89.1  \pm  2.7}{\percent}$   & $\SI{93.4  \pm  0.5}{\percent}$   & $\SI{84.7  \pm  4.7}{\percent}$   & $\SI{86.6  \pm 1.9}{\percent}$ \\
4&   $\SI{90.5  \pm  1.9}{\percent}$   & $\SI{92.9  \pm  0.4}{\percent}$   & $\SI{85.8  \pm  2.6}{\percent}$   & $\SI{86.1  \pm 1.4}{\percent}$ \\
5&   $\SI{90.5  \pm  2.3}{\percent}$   & $\SI{92.2  \pm  0.7}{\percent}$   & \cellcolor[RGB]{75,228,141}$\SI{88.4  \pm  1.5}{\percent}$   & $\SI{85.2  \pm 0.6}{\percent}$ \\
\hline
\multicolumn{1}{|c|}{} & \multicolumn{2}{c|}{3.5 inch} & \multicolumn{2}{c|}{4.5 inch} \\
\hline
\multicolumn{1}{|c|}{\makecell{Landscape \\ Number}} & Test Set &Training Set & Test Set& Training Set \\
\hline
1&   \cellcolor[RGB]{75,228,141}$\SI{92.2  \pm  5.4}{\percent}$   & $\SI{93.2  \pm  1.8}{\percent}$   & \cellcolor[RGB]{204,229,255}$\SI{66.8  \pm  5.6}{\percent}$   & $\SI{78.5  \pm 3.0}{\percent}$ \\
2&  $\SI{78.3  \pm  4.3}{\percent}$   & $\SI{85.2  \pm  2.1}{\percent}$   & \cellcolor[RGB]{204,229,255}$\SI{65.3  \pm  4.3}{\percent}$   & $\SI{69.3  \pm 2.3}{\percent}$ \\
3&  $\SI{82.6  \pm  4.3}{\percent}$   & $\SI{83.6  \pm  2.0}{\percent}$   & \cellcolor[RGB]{75,228,141} $\SI{68.6  \pm  4.8}{\percent}$   & $\SI{72.6  \pm 1.5}{\percent}$ \\
4&  $\SI{84.3  \pm  6.8}{\percent}$   & $\SI{84.3  \pm  2.8}{\percent}$   & \cellcolor[RGB]{204,229,255}$\SI{67.6  \pm  5.2}{\percent}$   & $\SI{71.8  \pm 2.6}{\percent}$ \\
5&  $\SI{85.7  \pm  4.8}{\percent}$   & $\SI{84.3  \pm  3.1}{\percent}$   & \cellcolor[RGB]{204,229,255}$\SI{66.8  \pm  5.5}{\percent}$   & $\SI{73.0  \pm 3.4}{\percent}$ \\
\hline
\end{tabular}
\end{table}

\begin{table}[h]
\centering
\caption{Persistence Landscape Results with Logistic Regression (LR) classifier. Landscape number column represents which landscapes was used to extract features.}
\label{tab:result_persistence_landscape-LR}
\begin{tabular}{|c|c|c|c|c|}
\hline
\multicolumn{1}{|c|}{} & \multicolumn{2}{c|}{2 inch} & \multicolumn{2}{c|}{2.5 inch} \\
\hline
\multicolumn{1}{|c|}{\makecell{Landscape \\ Number}} & Test Set &Training Set & Test Set& Training Set \\
\hline
1&    \cellcolor[RGB]{75,228,141}$\SI{95.7  \pm 1.6}{\percent}$   & $\SI{98.3  \pm 0.4}{\percent}$   & \cellcolor[RGB]{204,229,255}$\SI{82.6  \pm 4.2}{\percent}$   & $\SI{91.3  \pm 2.5}{\percent}$ \\
2&    $\SI{85.5  \pm 1.9}{\percent}$   & $\SI{93.7  \pm 1.0}{\percent}$   & \cellcolor[RGB]{204,229,255}$\SI{86.3  \pm 4.7}{\percent}$   & $\SI{91.5  \pm 2.1}{\percent}$ \\
3&    $\SI{88.5  \pm 1.1}{\percent}$   & $\SI{93.8  \pm 0.6}{\percent}$   & \cellcolor[RGB]{204,229,255}$\SI{84.7  \pm 4.3}{\percent}$   & $\SI{92.0  \pm 2.6}{\percent}$ \\
4&    $\SI{89.4  \pm 1.6}{\percent}$   & $\SI{94.8  \pm 0.8}{\percent}$   & \cellcolor[RGB]{75,228,141}$\SI{86.5  \pm 5.0}{\percent}$   & $\SI{90.8  \pm 2.4}{\percent}$ \\
5&    $\SI{88.3  \pm 1.7}{\percent}$   & $\SI{94.7  \pm 0.6}{\percent}$   & \cellcolor[RGB]{204,229,255}$\SI{86.0  \pm 3.4}{\percent}$   & $\SI{91.1  \pm 2.7}{\percent}$ \\
\hline
\multicolumn{1}{|c|}{} & \multicolumn{2}{c|}{3.5 inch} & \multicolumn{2}{c|}{4.5 inch} \\
\hline
\multicolumn{1}{|c|}{\makecell{Landscape \\ Number}} & Test Set &Training Set & Test Set& Training Set \\
\hline
1&    \cellcolor[RGB]{75,228,141}$\SI{91.3  \pm 3.4}{\percent}$   & $\SI{96.4  \pm 1.8}{\percent}$   & \cellcolor[RGB]{204,229,255}$\SI{63.1  \pm 3.5}{\percent}$   & $\SI{82.0  \pm 3.4}{\percent}$ \\
2&    $\SI{82.2  \pm 6.6}{\percent}$   & $\SI{90.7  \pm 2.6}{\percent}$   & \cellcolor[RGB]{204,229,255}$\SI{63.1  \pm 3.5}{\percent}$   & $\SI{82.0  \pm 3.4}{\percent}$ \\
3&    $\SI{87.8  \pm 6.4}{\percent}$   & $\SI{91.4  \pm 2.8}{\percent}$   & \cellcolor[RGB]{204,229,255}$\SI{64.2  \pm 4.5}{\percent}$   & $\SI{84.1  \pm 1.1}{\percent}$ \\
4&    \cellcolor[RGB]{204,229,255}$\SI{90.4  \pm 4.3}{\percent}$   & $\SI{91.8  \pm 1.5}{\percent}$   & \cellcolor[RGB]{204,229,255}$\SI{63.1  \pm 5.5}{\percent}$   & $\SI{82.9  \pm 2.4}{\percent}$ \\
5&    $\SI{85.7  \pm 4.8}{\percent}$   & $\SI{93.4  \pm 1.9}{\percent}$   & \cellcolor[RGB]{75,228,141}$\SI{65.9  \pm 4.6}{\percent}$   & $\SI{83.0  \pm 2.2}{\percent}$ \\
\hline
\end{tabular}
\end{table}

\begin{table}[h]
\centering
\caption{Persistence Landscape Results with Random Forest (RF) classifier. Landscape number column represents which landscapes was used to extract features.}
\label{tab:result_persistence_landscape-RF}
\begin{tabular}{|c|c|c|c|c|}
\hline
\multicolumn{1}{|c|}{} & \multicolumn{2}{c|}{2 inch} & \multicolumn{2}{c|}{2.5 inch} \\
\hline
\multicolumn{1}{|c|}{\makecell{Landscape \\ Number}} & Test Set &Training Set & Test Set& Training Set \\
\hline
1&    \cellcolor[RGB]{75,228,141}$\SI{96.1  \pm  1.7}{\percent}$    & $\SI{100.0 \pm  0.0}{\percent}$    & \cellcolor[RGB]{204,229,255}$\SI{86.7  \pm  5.6}{\percent}$    & $\SI{100.0 \pm  0.0}{\percent}$  \\
2&    $\SI{87.7  \pm  2.2}{\percent}$    & $\SI{100.0 \pm  0.0}{\percent}$    & \cellcolor[RGB]{75,228,141}$\SI{88.6  \pm  4.0}{\percent}$    & $\SI{100.0 \pm  0.0}{\percent}$  \\
3&    $\SI{88.0  \pm  1.9}{\percent}$    & $\SI{100.0 \pm  0.0}{\percent}$    & \cellcolor[RGB]{204,229,255}$\SI{87.4  \pm  2.4}{\percent}$    & $\SI{100.0 \pm  0.0}{\percent}$  \\
4&    $\SI{88.8  \pm  1.2}{\percent}$    & $\SI{100.0 \pm  0.0}{\percent}$    & \cellcolor[RGB]{204,229,255}$\SI{86.7  \pm  5.8}{\percent}$    & $\SI{100.0 \pm  0.0}{\percent}$  \\
5&    $\SI{89.6  \pm  2.2}{\percent}$    & $\SI{100.0 \pm  0.0}{\percent}$    & \cellcolor[RGB]{204,229,255}$\SI{86.7  \pm  5.8}{\percent}$    & $\SI{100.0 \pm  0.0}{\percent}$  \\
\hline
\multicolumn{1}{|c|}{} & \multicolumn{2}{c|}{3.5 inch} & \multicolumn{2}{c|}{4.5 inch} \\
\hline
\multicolumn{1}{|c|}{\makecell{Landscape \\ Number}} & Test Set &Training Set & Test Set& Training Set \\
\hline
1&    \cellcolor[RGB]{75,228,141}$\SI{91.3  \pm  3.9}{\percent}$    & $\SI{100.0 \pm  0.0}{\percent}$    & \cellcolor[RGB]{204,229,255}$\SI{65.1  \pm  7.7}{\percent}$    & $\SI{100.0 \pm  0.0}{\percent}$  \\
2&    $\SI{80.9  \pm  7.1}{\percent}$    & $\SI{100.0 \pm  0.0}{\percent}$    & \cellcolor[RGB]{75,228,141}$\SI{66.9  \pm  2.9}{\percent}$    & $\SI{100.0 \pm  0.0}{\percent}$  \\
3&    $\SI{83.0  \pm  6.3}{\percent}$    & $\SI{100.0 \pm  0.0}{\percent}$    & \cellcolor[RGB]{204,229,255}$\SI{66.8  \pm  4.9}{\percent}$    & $\SI{100.0 \pm  0.0}{\percent}$  \\
4&    $\SI{80.4  \pm  5.6}{\percent}$    & $\SI{100.0 \pm  0.0}{\percent}$    & \cellcolor[RGB]{204,229,255}$\SI{63.6  \pm  4.2}{\percent}$    & $\SI{100.0 \pm  0.0}{\percent}$  \\
5&    $\SI{85.2  \pm  5.9}{\percent}$    & $\SI{100.0 \pm  0.0}{\percent}$    & \cellcolor[RGB]{204,229,255}$\SI{64.7  \pm  7.1}{\percent}$    & $\SI{100.0 \pm  0.0}{\percent}$  \\
\hline
\end{tabular}
\end{table}

\begin{table}[h]
\centering
\caption{Persistence Images Results with pixel size = 0.1.}
\label{tab:result_persistence_images-0p1}
\begin{tabular}{|c|c|c|c|c|}
\hline
\multicolumn{1}{|c|}{} & \multicolumn{2}{c|}{2 inch} & \multicolumn{2}{c|}{2.5 inch} \\
\hline
\multicolumn{1}{|c|}{Classifier} & Test Set &Training Set & Test Set& Training Set \\
\hline
SVM&    $\SI{82.5  \pm 1.4}{\percent}$   & $\SI{82.7  \pm 0.6}{\percent}$   & $\SI{79.3  \pm 4.2}{\percent}$   & $\SI{81.4  \pm 2.9}{\percent}$ \\
LR &    $\SI{80.2  \pm 2.0}{\percent}$   & $\SI{82.4  \pm 0.9}{\percent}$   & $\SI{77.0  \pm 7.0}{\percent}$   & $\SI{82.4  \pm 4.5}{\percent}$ \\
RF &    \cellcolor[RGB]{75,228,141}$\SI{96.4  \pm 1.1}{\percent}$   & $\SI{100.0 \pm 0.0}{\percent}$   & \cellcolor[RGB]{75,228,141}$\SI{85.8  \pm 5.3}{\percent}$   & $\SI{100.0 \pm 0.0}{\percent}$ \\
GB &    \cellcolor[RGB]{204,229,255}$\SI{95.9  \pm 1.5}{\percent}$   & $\SI{100.0 \pm 0.0}{\percent}$   & \cellcolor[RGB]{204,229,255}$\SI{84.4  \pm 4.4}{\percent}$   & $\SI{100.0 \pm 0.0}{\percent}$ \\

\hline
\multicolumn{1}{|c|}{} & \multicolumn{2}{c|}{3.5 inch} & \multicolumn{2}{c|}{4.5 inch} \\
\hline
\multicolumn{1}{|c|}{Classifier} & Test Set &Training Set & Test Set& Training Set \\
\hline
SVM&    $\SI{82.6  \pm 5.5}{\percent}$   & $\SI{83.2  \pm 2.5}{\percent}$   & \cellcolor[RGB]{204,229,255}$\SI{66.9  \pm 5.8}{\percent}$   & $\SI{63.7  \pm 2.9}{\percent}$ \\
LR &    $\SI{82.6  \pm 5.1}{\percent}$   & $\SI{85.7  \pm 2.0}{\percent}$   & $\SI{62.7  \pm 5.9}{\percent}$   & $\SI{65.5  \pm 3.3}{\percent}$ \\
RF &    \cellcolor[RGB]{75,228,141}$\SI{93.0  \pm 2.9}{\percent}$   & $\SI{100.0 \pm 0.0}{\percent}$   & \cellcolor[RGB]{75,228,141}$\SI{72.5  \pm 6.4}{\percent}$   & $\SI{100.0 \pm 0.0}{\percent}$ \\
GB &    \cellcolor[RGB]{204,229,255}$\SI{91.3  \pm 4.3}{\percent}$   & $\SI{100.0 \pm 0.0}{\percent}$   & \cellcolor[RGB]{204,229,255}$\SI{68.5  \pm 2.8}{\percent}$   & $\SI{100.0 \pm 0.0}{\percent}$ \\
\hline
\end{tabular}
\end{table}

\begin{table}[h]
\centering
\caption{Persistence Images Results with pixel size = 0.05.}
\label{tab:result_persistence_images-0p05}
\begin{tabular}{|c|c|c|c|c|}
\hline
\multicolumn{1}{|c|}{} & \multicolumn{2}{c|}{2 inch} & \multicolumn{2}{c|}{2.5 inch} \\
\hline
\multicolumn{1}{|c|}{Classifier} & Test Set &Training Set & Test Set& Training Set \\
\hline
SVM&    $\SI{82.0  \pm 2.5}{\percent}$   & $\SI{83.4  \pm 1.2}{\percent}$   & $\SI{81.2  \pm 5.1}{\percent}$   & $\SI{80.2  \pm 2.8}{\percent}$ \\
LR &    $\SI{80.3  \pm 2.3}{\percent}$   & $\SI{79.5  \pm 1.0}{\percent}$   & $\SI{68.6  \pm 7.3}{\percent}$   & $\SI{75.2  \pm 3.4}{\percent}$ \\
RF &    \cellcolor[RGB]{75,228,141}$\SI{96.0  \pm 1.4}{\percent}$   & $\SI{100.0 \pm 0.0}{\percent}$   & \cellcolor[RGB]{75,228,141}$\SI{85.8  \pm 4.1}{\percent}$   & $\SI{100.0 \pm 0.0}{\percent}$ \\
GB &    \cellcolor[RGB]{204,229,255}$\SI{95.5  \pm 1.6}{\percent}$   & $\SI{100.0 \pm 0.0}{\percent}$   & \cellcolor[RGB]{204,229,255}$\SI{85.6  \pm 3.4}{\percent}$   & $\SI{100.0 \pm 0.0}{\percent}$ \\
\hline
\multicolumn{1}{|c|}{} & \multicolumn{2}{c|}{3.5 inch} & \multicolumn{2}{c|}{4.5 inch} \\
\hline
\multicolumn{1}{|c|}{Classifier} & Test Set &Training Set & Test Set& Training Set \\
\hline
SVM&    $\SI{85.2  \pm 4.0}{\percent}$   & $\SI{81.8  \pm 2.5}{\percent}$   & $\SI{63.4  \pm 5.2}{\percent}$   & $\SI{65.6  \pm 2.4}{\percent}$ \\
LR &    $\SI{80.9  \pm 5.9}{\percent}$   & $\SI{85.9  \pm 3.0}{\percent}$   & $\SI{63.6  \pm 4.5}{\percent}$   & $\SI{64.8  \pm 2.1}{\percent}$ \\
RF &    \cellcolor[RGB]{75,228,141}$\SI{90.9  \pm 3.0}{\percent}$   & $\SI{100.0 \pm 0.0}{\percent}$   & \cellcolor[RGB]{75,228,141}$\SI{70.7  \pm 3.1}{\percent}$   & $\SI{100.0 \pm 0.0}{\percent}$ \\
GB &    \cellcolor[RGB]{204,229,255}$\SI{90.4  \pm 4.7}{\percent}$   & $\SI{100.0 \pm 0.0}{\percent}$   & \cellcolor[RGB]{204,229,255}$\SI{70.0  \pm 5.4}{\percent}$   & $\SI{100.0 \pm 0.0}{\percent}$ \\
\hline
\end{tabular}
\end{table}

\begin{table}[h]
\centering
\caption{Template Function Results for $H_{0}$ diagrams.}
\label{tab:result_template_functions_H0}
\begin{tabular}{|c|c|c|c|c|}
\hline
\multicolumn{1}{|c|}{} & \multicolumn{2}{c|}{2 inch} & \multicolumn{2}{c|}{2.5 inch} \\
\hline
\multicolumn{1}{|c|}{Classifier} & Test Set &Training Set & Test Set& Training Set \\
\hline
SVM&    $\SI{82.4  \pm 1.8}{\percent}$   & $\SI{80.6  \pm 1.2}{\percent}$   & $\SI{82.1  \pm 6.5}{\percent}$   & $\SI{84.8  \pm 2.2}{\percent}$ \\
LR &    $\SI{81.7  \pm 2.1}{\percent}$   & $\SI{83.7  \pm 1.2}{\percent}$   & $\SI{78.1  \pm 7.4}{\percent}$   & $\SI{85.9  \pm 2.3}{\percent}$ \\
RF &    \cellcolor[RGB]{75,228,141}$\SI{87.6  \pm 4.0}{\percent}$   & $\SI{100.0 \pm 0.0}{\percent}$   & \cellcolor[RGB]{75,228,141}$\SI{86.5  \pm 2.5}{\percent}$   & $\SI{100.0 \pm 0.0}{\percent}$ \\
GB &    \cellcolor[RGB]{204,229,255}$\SI{86.7  \pm 2.5}{\percent}$   & $\SI{99.5  \pm 0.5}{\percent}$   & \cellcolor[RGB]{204,229,255}$\SI{84.9  \pm 5.7}{\percent}$   & $\SI{100.0 \pm 0.0}{\percent}$ \\

\hline
\multicolumn{1}{|c|}{} & \multicolumn{2}{c|}{3.5 inch} & \multicolumn{2}{c|}{4.5 inch} \\
\hline
\multicolumn{1}{|c|}{Classifier} & Test Set &Training Set & Test Set& Training Set \\
\hline
SVM&    \cellcolor[RGB]{75,228,141}$\SI{83.5  \pm 6.7}{\percent}$   & $\SI{81.4  \pm 3.5}{\percent}$   & $\SI{62.4  \pm 5.5}{\percent}$   & $\SI{66.0  \pm 2.8}{\percent}$ \\
LR &    $\SI{71.7  \pm 5.2}{\percent}$   & $\SI{90.0  \pm 1.8}{\percent}$   & \cellcolor[RGB]{204,229,255}$\SI{69.3  \pm 4.6}{\percent}$   & $\SI{70.9  \pm 2.2}{\percent}$ \\
RF &    \cellcolor[RGB]{204,229,255}$\SI{78.7  \pm 5.3}{\percent}$   & $\SI{100.0 \pm 0.0}{\percent}$   & \cellcolor[RGB]{75,228,141}$\SI{72.4  \pm 3.9}{\percent}$   & $\SI{100.0 \pm 0.0}{\percent}$ \\
GB &    \cellcolor[RGB]{204,229,255}$\SI{77.8  \pm 10.0}{\percent}$  & $\SI{100.0 \pm 0.0}{\percent}$   & \cellcolor[RGB]{204,229,255}$\SI{72.2  \pm 5.5}{\percent}$   & $\SI{100.0 \pm 0.0}{\percent}$ \\
\hline
\end{tabular}
\end{table}

\begin{table}[h]
\centering
\caption{Template Function Results for $H_{1}$ diagrams.}
\label{tab:result_template_functions_H1}
\begin{tabular}{|c|c|c|c|c|}
\hline
\multicolumn{1}{|c|}{} & \multicolumn{2}{c|}{2 inch} & \multicolumn{2}{c|}{2.5 inch} \\
\hline
\multicolumn{1}{|c|}{Classifier} & Test Set &Training Set & Test Set& Training Set \\
\hline
SVM&    $\SI{84.0  \pm 1.9}{\percent}$   & $\SI{85.9  \pm 1.5}{\percent}$   & $\SI{84.0  \pm 7.3}{\percent}$   & $\SI{86.4  \pm 2.9}{\percent}$ \\
LR &    $\SI{87.5  \pm 1.7}{\percent}$   & $\SI{90.1  \pm 0.7}{\percent}$   & $\SI{75.8  \pm 6.9}{\percent}$   & $\SI{92.9  \pm 2.3}{\percent}$ \\
RF &    $\SI{90.5  \pm 1.6}{\percent}$   & $\SI{100.0 \pm 0.0}{\percent}$   & \cellcolor[RGB]{75,228,141}$\SI{89.3  \pm 4.6}{\percent}$   & $\SI{100.0 \pm 0.0}{\percent}$ \\
GB &    \cellcolor[RGB]{75,228,141}$\SI{91.5  \pm 1.2}{\percent}$   & $\SI{100.0 \pm 0.0}{\percent}$   & $\cellcolor[RGB]{204,229,255}\SI{85.3  \pm 4.7}{\percent}$   & $\SI{100.0 \pm 0.0}{\percent}$ \\
\hline
\multicolumn{1}{|c|}{} & \multicolumn{2}{c|}{3.5 inch} & \multicolumn{2}{c|}{4.5 inch} \\
\hline
\multicolumn{1}{|c|}{Classifier} & Test Set &Training Set & Test Set& Training Set \\
\hline
SVM&    \cellcolor[RGB]{204,229,255}$\SI{83.0  \pm 4.9}{\percent}$   & $\SI{82.0  \pm 2.4}{\percent}$   & \cellcolor[RGB]{204,229,255}$\SI{64.7  \pm 3.6}{\percent}$   & $\SI{64.8  \pm 1.8}{\percent}$ \\
LR &    \cellcolor[RGB]{204,229,255}$\SI{78.7  \pm 10.0}{\percent}$  & $\SI{100.0 \pm 0.0}{\percent}$   & \cellcolor[RGB]{204,229,255}$\SI{64.1  \pm 4.4}{\percent}$   & $\SI{81.8  \pm 2.8}{\percent}$ \\
RF &    \cellcolor[RGB]{204,229,255}$\SI{81.3  \pm 8.5}{\percent}$   & $\SI{100.0 \pm 0.0}{\percent}$   & \cellcolor[RGB]{204,229,255}$\SI{61.4  \pm 5.9}{\percent}$   & $\SI{100.0 \pm 0.0}{\percent}$ \\
GB &    \cellcolor[RGB]{75,228,141}$\SI{83.9  \pm 5.5}{\percent}$   & $\SI{100.0 \pm 0.0}{\percent}$   & \cellcolor[RGB]{75,228,141}$\SI{65.1  \pm 6.1}{\percent}$   & $\SI{100.0 \pm 0.0}{\percent}$ \\
\hline
\end{tabular}
\end{table}

\begin{table}
\centering
\caption{Carlsson Coordinates Results.}
\label{tab:result_carlsson_coordinates}
\begin{tabular}{|c|c|c|c|c|}
\hline
\multicolumn{1}{|c|}{} & \multicolumn{2}{c|}{2 inch} & \multicolumn{2}{c|}{2.5 inch} \\
\hline
\multicolumn{1}{|c|}{Classifier} & Test Set &Training Set & Test Set& Training Set \\
\hline
SVM&$\SI{87.8\pm	2.0}{\percent}$&	$\SI{87.1\pm	1.2}{\percent}$&$\SI{72.1\pm	7.1}{\percent}$&	$\SI{79.7\pm	3.9}{\percent}$\\
LR &\cellcolor[RGB]{204,229,255}$\SI{93.1\pm	1.8}{\percent}$&	$\SI{92.9\pm	0.9}{\percent}$&\cellcolor[RGB]{75,228,141}$\SI{86.3\pm	6.2}{\percent}$&	$\SI{100.0\pm	0.0}{\percent}$\\
RF &\cellcolor[RGB]{204,229,255}$\SI{93.0\pm	2.0}{\percent}$&	$\SI{100.0\pm	0.0}{\percent}$&$\SI{69.5\pm	4.8}{\percent}$&	$\SI{70.9\pm	2.6}{\percent}$\\
GB &\cellcolor[RGB]{75,228,141}$\SI{93.6\pm	1.7}{\percent}$&	$\SI{100.0\pm	0.0}{\percent}$&\cellcolor[RGB]{204,229,255}$\SI{84.7\pm	5.0}{\percent}$&	$\SI{100.0\pm	0.0}{\percent}$\\

\hline
\multicolumn{1}{|c|}{} & \multicolumn{2}{c|}{3.5 inch} & \multicolumn{2}{c|}{4.5 inch} \\
\hline
\multicolumn{1}{|c|}{Classifier} & Test Set &Training Set & Test Set& Training Set \\
\hline
SVM&\cellcolor[RGB]{204,229,255}$\SI{95.2\pm	4.5}{\percent}$&	$\SI{95.2\pm	1.9}{\percent}$&\cellcolor[RGB]{204,229,255}$\SI{68.1\pm	6.5}{\percent}$&	$\SI{68.6\pm	2.4}{\percent}$\\
LR &$\SI{90.9\pm	9.2}{\percent}$&	$\SI{93.6\pm	1.4}{\percent}$&\cellcolor[RGB]{204,229,255}$\SI{70.8\pm	4.0}{\percent}$&	$\SI{72.6\pm	3.0}{\percent}$\\
RF &\cellcolor[RGB]{75,228,141}$\SI{95.7\pm	4.3}{\percent}$&    $\SI{100.0\pm	0.0}{\percent}$&\cellcolor[RGB]{75,228,141}$\SI{72.2\pm	4.4}{\percent}$&	$\SI{100.0\pm	0.0}{\percent}$\\
GB &\cellcolor[RGB]{204,229,255}$\SI{92.2\pm	4.3}{\percent}$&	$\SI{100.0\pm	0.0}{\percent}$&\cellcolor[RGB]{204,229,255}$\SI{71.9\pm	3.3}{\percent}$&	$\SI{100.0\pm	0.0}{\percent}$\\
\hline
\end{tabular}
\end{table}

\begin{table}[h]
\centering
\begin{threeparttable}
\caption{Kernel Method Results with LibSVM package}
\label{tab:result_kernel_method}
\begin{tabular}{|c|c|c|c|c|c|}
\hline
\multicolumn{1}{|c|}{} & 2 inch &2.5 inch & 3.5 inch & 4.5 inch \\
\hline
\multicolumn{1}{|c|}{Kernel Scale ($\sigma$)} &\multicolumn{4}{c|}{Test Set Score and Deviation} \\
\hline
0.2  &  *       								& *       											  & $\SI{30.4 \pm 6.2}{\percent}$ & *  									        \\
0.25 &  \cellcolor[RGB]{75,228,141}$\SI{74.5}{\percent}$**    & \cellcolor[RGB]{75,228,141}$\SI{58.9 \pm 28.5}{\percent}$  & \cellcolor[RGB]{75,228,141}$\SI{87.0 \pm 3.6}{\percent}$ & \cellcolor[RGB]{75,228,141}$\SI{59.3 \pm 9.6}{\percent}$ \\
\hline
\end{tabular}
\begin{tablenotes}
\small
\item *These results are not available due to high computational time. \\
\item **This result belongs to first iteration for 2 inch overhang case. 
\end{tablenotes}
\end{threeparttable}
\end{table}


\begin{table}[h]
\centering
\caption{Path signature method results obtained with SVM classifier. Landscape number represent which landscapes was used to compute signatures.}
\label{tab:result_path_signature}
\begin{tabular}{|c|c|c|c|c|}
\hline
\multicolumn{1}{|c|}{} & \multicolumn{2}{c|}{2 inch} & \multicolumn{2}{c|}{2.5 inch} \\
\hline
\multicolumn{1}{|c|}{\makecell{Landscape \\ Number}} & Test Set &Training Set & Test Set& Training Set \\
\hline
1&	\cellcolor[RGB]{75,228,141}$\SI{83.0  \pm 2.9}{\percent}$   & 	$\SI{84.8  \pm 0.9}{\percent}$   & $\SI{82.7  \pm 5.3}{\percent}$   & 	$\SI{82.9  \pm 2.0}{\percent}$ \\
2&	$\SI{79.2  \pm 3.1}{\percent}$   & 	$\SI{80.2  \pm 1.0}{\percent}$   & 	\cellcolor[RGB]{75,228,141}$\SI{84.2  \pm 3.5}{\percent}$   & 	$\SI{80.4  \pm 1.4}{\percent}$ \\
3&	$\SI{78.6  \pm 1.8}{\percent}$   & 	$\SI{79.4  \pm 0.7}{\percent}$   & 	$\SI{79.1  \pm 2.9}{\percent}$   & 	$\SI{80.8  \pm 1.1}{\percent}$ \\
4&	$\SI{79.3  \pm 2.5}{\percent}$   & 	$\SI{79.5  \pm 0.8}{\percent}$   & 	$\SI{80.6  \pm 5.6}{\percent}$   & 	$\SI{78.6  \pm 2.3}{\percent}$ \\
5&	$\SI{0.0   \pm 0.0}{\percent}$   & 	$\SI{0.0   \pm 0.0}{\percent}$   & 	$\SI{80.9  \pm 5.1}{\percent}$   & 	$\SI{78.8  \pm 1.2}{\percent}$ \\
\hline
\multicolumn{1}{|c|}{} & \multicolumn{2}{c|}{3.5 inch} & \multicolumn{2}{c|}{4.5 inch} \\
\hline
\multicolumn{1}{|c|}{\makecell{Landscape \\ Number}} & Test Set &Training Set & Test Set& Training Set \\
\hline
1	&$\SI{81.2  \pm 6.3}{\percent}$   & $\SI{82.6  \pm 2.2}{\percent}$   & \cellcolor[RGB]{75,228,141}$\SI{70.0  \pm 6.4}{\percent}$   & $\SI{71.4  \pm 2.2}{\percent}$ \\
2	&$\SI{82.9  \pm 7.2}{\percent}$   & $\SI{81.8  \pm 2.4}{\percent}$   & $\SI{64.1  \pm 5.4}{\percent}$   & $\SI{67.0  \pm 2.4}{\percent}$ \\
3	&$\SI{81.2  \pm 10.5}{\percent}$  & $\SI{82.2  \pm 3.8}{\percent}$   & $\SI{67.7  \pm 6.6}{\percent}$   & $\SI{65.8  \pm 2.3}{\percent}$ \\
4	&\cellcolor[RGB]{75,228,141}$\SI{85.9  \pm 4.7}{\percent}$   & $\SI{80.8  \pm 1.6}{\percent}$   & $\SI{64.5  \pm 4.2}{\percent}$   & $\SI{65.4  \pm 1.2}{\percent}$ \\
5	&$\SI{82.4  \pm 9.1}{\percent}$   & $\SI{82.0  \pm 3.1}{\percent}$   & $\SI{64.8  \pm 6.9}{\percent}$   & $\SI{65.4  \pm 2.1}{\percent}$ \\
\hline
\end{tabular}
\end{table}